\documentclass[fleqn,usenatbib]{mnras}

\usepackage{newtxtext,newtxmath}

\usepackage[T1]{fontenc}
\usepackage{ae,aecompl}

\usepackage{graphicx}	
\usepackage{amsmath}	
\usepackage{amsmath, epsfig,natbib}
\usepackage{multicol}
\usepackage{color,ulem}
\usepackage{newtxtext,newtxmath}
\definecolor{webgreen}{rgb}{0,.5,0}
\definecolor{webbrown}{rgb}{.6,0,0}
\usepackage{subfigure}

\newcommand{\kms}{\mbox{$\>{\rm km\, s^{-1}}$}}
\newcommand{\pc}{\>{\rm pc}}
\newcommand{\kpc}{\mbox{$\>{\rm kpc}$}} 
\newcommand{\kkms}{\mbox{$\>{\rm kpc\, km\, s^{-1}}$}}
\newcommand{\kmsk}{\mbox{$\>{\rm km\, s^{-1}\, kpc^{-1}}$}}
\newcommand{\Gyr}{\mbox{$\>{\rm Gyr}$}}

\newcommand\degrees{^\circ}
\newcommand\gaia{{\it Gaia}}
\newcommand\roadmapping{{\it RoadMapping}}
\newcommand{\avg}[1]{\mbox{$\left<{#1}\right>$}}

\title [Action-based modelling in the presence of a bar ] 
{Quantifying the influence of bars on action-based dynamical modelling of disc galaxies}

\author[Ghosh et al.]
	{Soumavo Ghosh,$^{1}$\thanks{E-mail: ghosh@mpia-hd.mpg.de}
	 Wilma H. Trick, $^{2}$
	 Gregory M. Green $^{1}$ \\
$^{1}$ Max-Planck-Institut f\"{u}r Astronomie, K\"{o}nigstuhl 17, D-69117 Heidelberg, Germany\\
$^2$  Max-Planck-Institut für Astrophysik, Karl-Schwarzschild-Str. 1, 85748 Garching b. München, Germany\\}

 \date{Accepted 2023 May 16. Received 2023 April 28; in original form 2022 December 8}

\pubyear{2022}

\begin{document}
\label{firstpage}
\pagerange{\pageref{firstpage}--\pageref{lastpage}}
\maketitle
\begin{abstract} 
Action-based dynamical modelling, using stars as dynamical tracers, is an excellent diagnostic to estimate the underlying axisymmetric matter distribution of the Milky Way. However, the Milky Way's bar causes non-axisymmetric resonance features in the stellar disc. Using \roadmapping\ (an action-based dynamical modelling framework to estimate the gravitational potential and the stellar distribution function), we systematically quantify the robustness of action-based modelling in the presence of a bar. We construct a set of test-particle simulations of barred galaxies (with varying bar properties), and apply \roadmapping\ to different survey volumes (with varying azimuthal position, size) drawn from these barred  models. For realistic bar parameters, the global potential parameters are still recovered to within ${\sim \! 1 \! - \! 17}$ percent. However, with increasing bar strength, the best-fit values of the parameters progressively deviate from their true values. This happens due to a combination of radial heating, radial migration, and resonance overlap phenomena in our bar models. Furthermore, the azimuthal location and the size of the survey volumes play important roles in the successful recovery of the parameters. Survey volumes along the bar major axis produce larger (relative) errors in the best-fit parameter values. In addition, the potential parameters are better recovered for survey volumes with larger spatial coverage. As the Sun is located just ${\sim \! 28 \! - \! 33 \degrees}$ behind the bar's major axis, an estimate for the bar-induced systematic bias -- as provided by this study -- is therefore crucial for future modelling attempts of the Milky Way.

\end{abstract}


\section{Introduction}
\label{sec:Intro}
One of the key objectives in the field of Milky Way (MW) dynamics is to constrain the underlying gravitational potential, which reflects the spatial distribution of matter (both the baryons and the dark matter). The second \gaia \ Data Release (hereafter \gaia\ DR2) has revolutionised the study of Galactic archaeology and dynamics by providing 6-D position-velocity measurements for millions of stars in the Solar Neighbourhood (and beyond) with an unprecedented precision \citep{Katzetal2018}. Furthermore, the recent third \gaia \ Data Release (hereafter \gaia\ DR3) has increased the number of stars with rich 6-D position-velocity information by a factor of $\sim$ five \citep[e.g.][]{GaiaCollaboration2022}. The challenge remains: with this rich phase-space information of several tens of millions stars in hand, can we successfully constrain the underlying potential of the MW?
\par
In principle, a \textit{direct} measurement of an individual star's acceleration would provide information on the local gradient of the gravitational potential of the MW \citep[e.g.,][]{BT08,Quercellinietal2008,Ravietal2019,Chakrabartietal2020}. However, the scale of these accelerations, on the order of $1$~cm~s$^{-1}$~yr$^{-1}$, is well below the precision achieved by ongoing spectroscopic and astrometric instruments \citep{silverwood_easther_2019}. To date, only a handful of acceleration measurements have been made at this level of accuracy \citep{Chakrabartietal2021,Phillipsetal2021,Chakrabartietal2022}. This leaves us with only the means of inferring the MW's gravitational potential: by employing rigorous dynamical modelling of the kinematics of baryons \citep[e.g. stars, gas; see][]{BT08,BinneyandMcMillan2011,RixandBovy2013}.
\par
Past theoretical efforts to recover the gravitational potential of the MW have taken various approaches towards dynamical modelling of the discrete collisionless tracers -- namely, the stars. To mention a few, the underlying potential can be determined via Jeans modelling \citep{KuijkenandGilmore1989,BovyandTremaine2012,Garbarietal2012,Zhangetal2013,Buedenbenderetal2015}, torus modelling \citep{McMillanandBinney2008,McMillanandBinney2012,McMillanandBinney2013}, action-based distribution function (DF) modelling -- both with parametric DFs \citep{BovyandRix2013,Piffletal2014,SandersandBinney2015,DasandBinney2016,Tricketal2016,Tricketal2017} as well as with marginalization over non-parametric DFs \citep{Magorrian2014}, and made-to-measure modelling \citep{SyerandTremaine1996,deLorenzietal2007,HuntandKawata2014}. Most of these techniques assume a DF (implicitly or explicitly) to capture the distribution of the stars, and these assumed DFs are often simple parametric models. Therefore, the \textit{successful} recovery of the underlying potential relies on the accuracy or robustness of the assumed DF model. A few recent works have proposed methods for recovering the underlying potential that do not depend on simplified analytic DFs \citep[e.g., see][]{Greenetal2020,Anetal2021,Naiketal2022,Buckleyetal2022,Greenetal2022}, which thus are well adapted to systems with rich and complex phase-space structures, such as the MW \citep[e.g., see][]{Antojaetal2018,Tricketal2019}. Nevertheless, these methods still assume that the MW is stationary.
\par
However, the MW harbours a central stellar bar \citep[e.g.,][]{LisztandBurton1980,Binneyetal1991,Weinberg1992,Binneyetal1997,BlitzandSpergel1991,Hammersleyetal2000,WegandGerhard2013} and also contains spiral structure in the disc region \citep[e.g.,][see also \citealt{Valle2005,Valle2008}]{Oort1958,GeorgelinandGeorgelin1976,Gerhard2002,Churchweletal2009,Reidetal2014}, an $m=1$ lopsidedness, and warps in the outer disc region \citep[e.g.,][]{KalberlaandDedes2008,Romeroetal2019}. These non-axisymmetric structures can induce non-circular motions \citep[e.g.,][]{Randriamampandryetal2016}, heating of the stellar distribution function \citep{Sahaetal2010,Grandetal2016,Pinnaetal2018,Ghoshetal2022a}, radial mixing of the stars \citep[e.g.,][]{SellwoodBinney2002,Roskaretal2008,SchonrichBinney2009,Roskaretal2012}, large-scale vertical breathing motions \citep{Debattista2014,Monarietal2015,Ghoshetal2022,Kumaretal2022,Khachaturyantsetal2022}, and ridges in the ${(V_{\phi},R)}$-plane \citep{Fragkoudietal2019}. Furthermore, in presence of non-axisymmetric structures (e.g., bars and spirals), the actions of individual stars no longer remain \textit{constants of motion} \citep[e.g.,][]{BT08,Solwayetal2012,Vera_Ciro_2016}, violating one of the central assumptions of action-based dynamical modelling of the (steady-state) MW. Therefore, in order to interpret the results of action-based modelling of the MW, it is crucial to quantify the dynamical role of these asymmetries on the recovery of the underlying axisymmetric potential. \citet{Tricketal2017} investigated the dynamical role of the spiral arms in the \roadmapping\ framework \citep[an action-based dynamical modelling framework of the MW; see][]{Tricketal2016}. Firstly, they demonstrated that if the data coverage of the Galactic disc (i.e. the survey volume) is large enough (${\sim \! 4 \! - \! 5 \kpc}$ around the Sun), \roadmapping\ can successfully interpolate over the spiral arms and recover the \textit{global} potential parameters. Secondly, they showed that in the case of small survey volumes (${\sim \! 500 \pc}$ around the Sun), \roadmapping\ still provides a reliable measurement of \textit{local} gravitational forces, which for inter-arm regions are also fairly close to the \textit{global} axisymmetric gravitational forces. However, such a quantitative study for the stellar bar is still missing to date.
\par
In this paper, we systematically investigate the dynamical influence of a stellar bar on action-based dynamical modelling of a disc galaxy, by using a set of test particle models of barred galaxies with varying bar properties (e.g., strength, pattern speed). For this work, we continue to make use of the \roadmapping\ (``Recovery of the Orbit Action Distribution of Mono-Abundance Populations and Potential INference
for our Galaxy'') framework to quantify the effect of stellar bars in recovering the parameters of the underlying axisymmetric potential. \\
\par
The rest of the paper is organized as follows.
Section~\ref{sec:sim_setup} provides the details of the initial axisymmetric, equilibrium test-particle simulation set-up, while section~\ref{sec:bar_model} provides the detailed properties of our barred models. Section~\ref{sec:roadmapping} contains a brief overview of the \roadmapping\ framework, which we use later in this paper.  Section~\ref{sec:results} presents our findings on the dynamical effect of stellar bars in the recovery of the underlying axisymmetric potential and the DF.  Section~\ref{sec:discussion} discusses caveats and implications of this work for dynamical modeling of the Milky Way, while section \ref{sec:summary} summarizes the main findings of this work.

\section{The axisymmetric equilibrium galaxy model}
\label{sec:sim_setup}
Here, we describe the details of our equilibrium test particle model for a disc galaxy. The following potential and distribution function model are used both for generating the initial snapshot of the test particle simulation, and as model in the \roadmapping\ framework to fit the data. Throughout this work, we make extensive use of the {\sc{galpy}} Python library by \citet{Bovy2015}.\footnote{The full {\sc{galpy}} package can be downloaded from \href{https://github.com/jobovy/galpy} {https://github.com/jobovy/galpy}.}

\subsection{Gravitational potential}   
\label{sec:gravpot}
Our initial, axisymmetric equilibrium model of the Galactic potential $\Phi_{\rm axi}(R,z)$ consists of a spherical bulge, a stellar disc, and a spherical dark matter (DM) halo. We model the stellar disc potential as a Miyamoto-Nagai disc \citep{MiyamatoandNagai1975}, which has the form
\begin{equation}
\Phi_{\rm disc}(R,z) = -\frac{GM_{\rm disc}}{\sqrt{R^2+\left(a_{\rm disc}+\sqrt{z^2+b_{\rm disc}^2}\right)^2}}\,,
\label{eq:pot_disc}
\end{equation}
\noindent where $a_{\rm disc}$ and $b_{\rm disc}$ control the disc scale length and the scale height, respectively. The bulge is modelled with a Hernquist density profile \citep{Hernquist1990} of the form
\begin{equation}
    \rho_{\rm bul}(r)=\frac{M_{\rm bul}}{2\pi a_{\rm bul}^3}\frac{1}{(r/a_{\rm bul})(1+r/a_{\rm bul})^{3}}\,,
    \label{eq:pot_bulge}
\end{equation}
\noindent where $M_{\rm bul}$ and $a_{\rm bul}$ represent the total bulge mass and the bulge scale radius, respectively. Here, $r$ and $R$ are the radius in the spherical and the cylindrical coordinates, respectively.  In {\sc{galpy}}, instead of directly setting the parameter $M_{\rm bul}$, we set the relative bulge-to-total contribution to the radial force at the `Solar' location, $R_{\odot}$ ($8 \kpc$ here, kept fixed throughout) which is defined as 
\begin{equation}
f_{\rm bulge} = \left( \frac{F_{R, \rm bulge}}{F_{R, \rm total}} \right)_{R= R_{\odot}, z =0}\,.
\label{eq:relative_bulgeradialforce}
\end{equation}
\par
The DM halo is  modelled with a NFW profile  of the form \citep{NFW1996}
\begin{equation}
    \rho_{\rm DM}(r)=\frac{M_{\rm DM}}{4\pi a_{\rm halo}^3}\frac{1}{(r/a_{\rm halo})(1+r/a_{\rm halo})^2}\,,
    \label{eq:pot_halo}
\end{equation}
\noindent where $M_{\rm DM}$ and $a_{\rm halo}$ are the total mass and the scale radius of the DM halo, respectively. The relative halo-to-disc contribution to the radial force at the `Solar' location, given by 
\begin{equation}
f_{\rm halo} = \left( \frac{F_{R, \rm halo}}{F_{R, \rm disc}+F_{R, \rm halo}} \right)_{R= R_{\odot}, z =0}\,,
\label{eq:relative_radialforce}
\end{equation}
\noindent is used as a free parameter in our model to indirectly set $M_{\rm DM}$. Similarly, we use the circular velocity at the `Solar' location, defined as 
\begin{equation}
v_{\rm circ} (R \! = \! R_{\odot}) = \sqrt{R \frac{\partial \Phi_{\rm axi}}{\partial R}} \, \Bigg |_{R= R_{\odot}, z =0}\,,
\end{equation}
\noindent as another free parameter to model the total mass of the galactic model. The values of the parameters of the potential $\Phi_{\rm axi}$, used for the initial model, are listed in Table~\ref{tab:equilibrium_params}. Based on our still uncertain knowledge of the MW's potential, especially beyond the Solar neighbourhood, we chose parameters that are roughly plausible for the MW \citep{2016ARA&A..54..529B}, but do not claim this base model to be an exact representation of the MW.

\begin{table}
\caption{Parameters for the initial equilibrium MW model.}
\begin{tabular}{ccc}
\hline
\hline
\multicolumn{3}{c}{\textbf{Potential parameters ($p_{\Phi}$)}} \\
\hline
`Solar' Galactocentric radius ($R_{\odot}$) & : & $8 \kpc$\\
circular velocity at  $R_{\odot}$ ($v_{\rm circ}(R \! = \! R_{\odot})$) & : & $220 \kms$\\
Disc scale length ($a_{\rm disc}$) & : & $2.5 \kpc$\\
Disc scale height ($b_{\rm disc}$) & : & $0.3 \kpc$\\
Bulge scale length ($a_{\rm bul}$) & : & $0.6 \kpc$\\
Relative bulge-to-total contribution ($f_{\rm bulge}$) & : & 0.05\\
DM halo scale length ($a_{\rm halo}$) & : & $18 \kpc$\\
Relative halo-to-disc contribution ($f_{\rm halo}$) & : & 0.3\\
\hline
\multicolumn{3}{c}{\textbf{DF parameters ($p_{\rm DF}$)}} \\
\hline
Radial tracer density scale length ($h_R$) & : & $2.5 \kpc$\\
Radial velocity dispersion at $R_{\odot}$ ($\sigma_{R,0}$) & : & $33 \kms$\\
Vertical velocity dispersion at $R_{\odot}$ ($\sigma_{z,0}$) & : & $25 \kms$\\
Radial velocity dispersion scale length ($h_{\sigma, R}$) & : & $8 \kpc$ \\
Vertical velocity dispersion scale length ($h_{\sigma, z}$) & : & $7 \kpc$ \\
\hline
\hline
\end{tabular}
\label{tab:equilibrium_params}
\end{table}

\subsection{Stellar disc distribution function}
\label{sec:stellar_df}
In order to create an equilibrium stellar disc model in this assumed combined disc, bulge and halo potential, we make use of the \textit{quasi-isothermal} DF (qDF), which is an action-based orbit DF, introduced by \citet{Binney2010} and \citet{BinneyandMcMillan2011}. As the actions $\vec{J}\equiv(J_R,L_z,J_z)$ are integrals of motion in an unperturbed axisymmetric potential, the qDF satisfies the steady-state Collisionless Boltzman equation (CBE) by construction. $J_R$, $J_{\phi}$ (i.e., angular momentum $L_z$), and $J_z$ are the radial, azimuthal, and vertical actions, respectively. The functional form of the qDF is given by
\begin{equation}
{\rm {qDF}}({\bf{J}}| p_{\rm DF})
  = f_{\sigma_R} (J_R, L_z | p_{\rm DF})
    \times f_{\sigma_z} (J_z, L_z | p_{\rm DF}) \,,
\label{eq:qDF}
\end{equation}
\noindent where
\begin{equation}
p_{\rm DF} \equiv \{ h_R, \sigma_{R,0},  \sigma_{z,0}, h_{\sigma,R}, h_{\sigma,z} \}
\label{eq:params_to_fit}
\end{equation}
\noindent are the free parameters of the qDF. For our true base model, we set them to the values listed in Table~\ref{tab:equilibrium_params}. They roughly mimic a high-[Fe/H], low-[$\alpha/$Fe], thin disc component with velocity dispersions close to those at the Solar location \citep{Bovyetal2012,BovyandRix2013,Tricketal2019}. The functions $f_{\sigma_R}$ and $f_{\sigma_z}$ govern the action distributions for $J_R$ and $J_z$, as function of the guiding-centre radius $R_g(L_z)$. They are related to the velocity dispersion profiles $\sigma_R(R)$ and $\sigma_z(R)$, and are given by
\begin{equation}
\begin{split}
f_{\sigma_R} (J_R, L_z | p_{\rm DF})
  = n (R_{\rm g}) \times \frac{\Omega}{\pi \kappa \sigma_R^2 (R_{\rm g})}
    \exp \left[ - \frac{\kappa J_R}{\sigma_R^2 (R_{\rm g})} \right] \\
    \times \, [1+ \tanh(L_z/L_0)]\,,
\end{split}
\end{equation}
\noindent and
\begin{equation}
\begin{split}
f_{\sigma_z} (J_z, L_z | p_{\rm DF})
  = \frac{\nu}{2 \pi \sigma_z^2 (R_{\rm g})}
    \exp \left[- \frac{\nu J_z}{\sigma_z^2(R_{\rm g})} \right]\,.
\end{split}
\end{equation}
\noindent Here, $\Omega$, $\kappa$, and $\nu$ are the circular, radial (epicyclic), and vertical frequencies, respectively. Furthermore, $n (R_{\rm g})$ is the stellar radial density.  Following \citet{BinneyandMcMillan2011}, we assume the following scaling relations to mimic an exponential stellar disc:
\begin{align}
n (R_{\rm g} | p_{\rm DF}) &\propto
  \exp \left[-\frac{R_{\rm g}}{h_R} \right],
  \\
\sigma_R (R_{\rm g} | p_{\rm DF}) &\propto
  \sigma_{R,0} \times \exp \left[-\frac{R_{\rm g} - R_{\odot}}{h_{\sigma,R}} \right],
  \\
\sigma_z (R_{\rm g} | p_{\rm DF}) &\propto
  \sigma_{z,0} \times \exp \left[-\frac{R_{\rm g} - R_{\odot}}{h_{\sigma,z}} \right].
\end{align}
\noindent The same scaling relations were also used previously in \citet{RixandBovy2013,Tricketal2016, Tricketal2017}. We sample model stars from the qDF with the parameters given in Table~\ref{tab:equilibrium_params} to generate the initial snapshot of our test particle simulation.
 In the \roadmapping\ analysis, we fit a subset of the DF parameters, $\{ \ln h_R, \ln\sigma_{R,0},  \ln \sigma_{z,0} \} \subset p_{\rm DF}$, holding all others fixed to their true values in Table~\ref{tab:equilibrium_params}.
 
 \subsection{Generating a test particle disc}
\label{sec:generate_equi_model}
One pivotal step to (i) create the initial axisymmetric model and to (ii) run the \roadmapping\ modelling (see section~\ref{sec:roadmapping}) is to convert the action-space DF to a 3D spatial density, which requires an assumed potential. For the assumed qDF$({\bf J} \, | \, p_{\rm DF}, p_{\Phi})$ (see Eq.~\ref{eq:qDF}), the corresponding density, $\rho_{\rm DF} ({\bf x} \, | \, p_{\rm DF}, p_{\Phi})$ is calculated as follows
\begin{equation}
\rho_{\rm DF} ({\bf x} \, | \, p_{\rm DF}, p_{\Phi})
  = \int_{-\infty}^{\infty} {\rm qDF}({\bf J} \, | \, p_{\rm DF}, p_{\Phi}) d{\bf v}
  \,,
\end{equation}
\noindent which, when written in terms of the cylindrical coordinates $(R, \phi, z)$, reduces to 
\begin{equation}
\rho_{\rm DF} (R, |z| \, | \, p_{\rm DF}, p_{\Phi})
  = \int_{-\infty}^{\infty}
    {\rm qDF}({\bf J}[R, z, {\bf v} | p_{\Phi}] \, | \, p_{\rm DF})
    d^3 v \,.
\label{eq:density_normalisation}
\end{equation}
\citet{Tricketal2016} details how this integral can be evaluated (using reasonable finite limits on the integrals) on an ${N_x \times N_x}$ regular density grid in the ${(R, z)}$-plane and subsequently interpolated. Following the terminology of \citet{Tricketal2016}, we use a relatively fine numerical grid, ${N_x = 20}$. For the velocity integrals, we use ${N_v = 28}$ and ${n_{\sigma} = 5.5}$. To set up the density interpolation grid, a total of ${N_x^2 \times N_v^3}$ actions need to be calculated \citep[see][]{BovyandRix2013, Tricketal2016}. For the assumed potential, the actions $\bf{J}$ are computed from positions $\vec{x}$ and velocities $\vec{v}$ using the \textit{St\"{a}ckel fudge} algorithm \citep{Binney2012} with a fixed focal length $\Delta = 0.45$. To speed up the action calculation, we use an interpolation grid for the actions as well \citep{Binney2012, Bovy2015}. 
\par
\begin{figure*}
\includegraphics[width=0.9\linewidth]{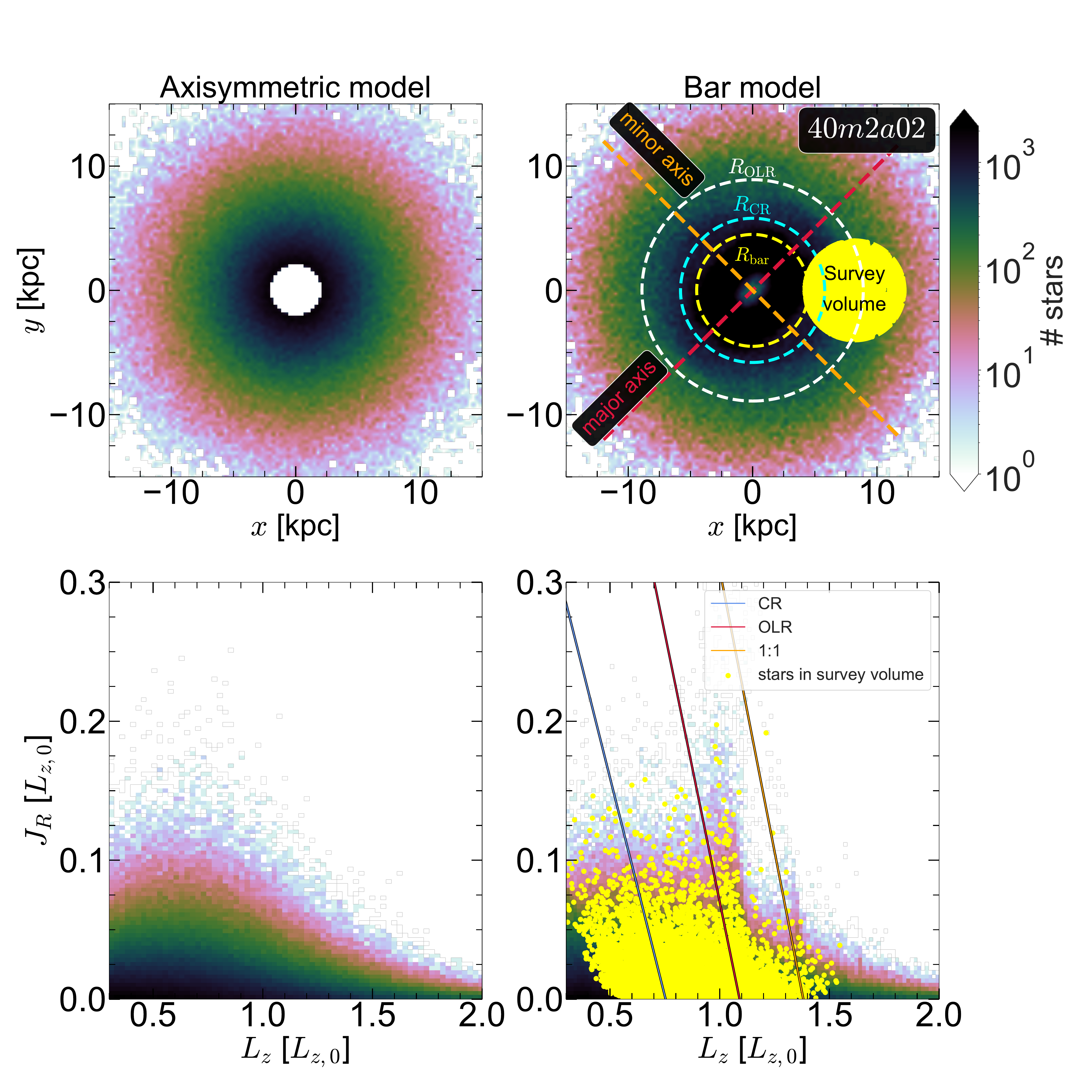}
\caption{
 Density distribution of stars in the test particle simulation, projected onto the ($x,y$)-plane (upper panels) and the ($J_R, L_z$)-plane (bottom panels) for the initial axisymmetric galaxy disc model (left column) and also one of the models affected by an intermediate-strength bar, \textbf{40m2a02}, integrated for $\sim 8.4 \Gyr$ (right column, for details see section~\ref{sec:bar_model}).
In the top-right panel, the yellow circular region denotes one such selected survey volume on which we perform the \roadmapping\ analysis (shown here is the \textbf{40m2a02Inter} dataset). The inner yellow dashed circle denotes the extent of bar ($R_{\rm bar} = 4.5 \kpc$), while the cyan and the white dashed circles denote the CR ($R_{\rm CR} = 5.8 \kpc$) and the OLR ($R_{\rm OLR} = 8.9 \kpc$) of the bar, respectively. The red and the orange dashed straight lines denote the bar major and the minor axes, respectively. In the bottom right panel, the yellow filled circles denote the selected stars whereas the straight lines denote the corresponding ARLs, for details see section~\ref{sec:recap_resonances_action_space}. As throughout this work, ${L_{z,0} = 8 \times 220 \kkms}$.
}
\label{fig:densmap_initialandfinal}
\end{figure*}

 The qDF and the corresponding density in Eq.~\ref{eq:density_normalisation} are used to sample a total of $5 \times 10^5$  particles, which constitute our initial axisymmetric stellar disc. The left panels of Fig.~\ref{fig:densmap_initialandfinal} show the corresponding density distribution of stars at ${t = 0}$ in a face-on projection (\textit{i.e.}, the ${(x,y)}$-plane) as well as in an action-space projection (the ${(J_R, L_z)}$-plane).
 
\section{The barred galaxy model}
\label{sec:bar_model}

Here, we briefly discuss how the bar models are generated and also how they compare with what is known about the MW bar. 

\subsection{The bar of the Milky Way}
\label{sec:mw_bar}
%
Here, we briefly summarize the properties (known to date) of the MW bar. 
The jury is still out on the exact pattern speed of the bar. This is important, as the bar pattern speed, together with the rotation curve, set the locations of the resonances (CR, ILR, OLR), which in turn, determine whether the Solar neighbourhood lies close to any of these resonances. Earlier bar pattern speed measurements favoured a fast-short bar scenario with a somewhat larger value of the pattern speed, ranging from $\sim \!\! 50$ to $60 \kmsk$ \citep[e.g., see][]{Fux1999,Dehnen2000,Debattistaetal2002,Bissantzetal2003,2014A&A...563A..60A}. However, more recent observational measurements favour a relatively lower value for the bar pattern speed, mostly converging towards $\sim \!\! 40 \kmsk$ \citep[e.g., see][]{Sormanietal2015,Lietal2016,Portailetal2017,2019MNRAS.490.4740B,Sandersetal2019,Clarkeetal2022,Lietal2022,Luceyetal2022}. 
Additionally, the MW bar is oriented at $\sim \! 28 \! - \! 33 \degrees$ with respect to the Solar neighbourhood \citep[e.g.,][]{Weggetal2015}.
\par
The bar strength and shape are even less well known. \citet{Portailetal2017} applied made-to-measure modelling to data in the bar region. Their slow-bar model was used by \citet{2017ApJ...840L...2P} and \citet{2019A&A...626A..41M} to compare the strength of the bar resonances in the Solar neighbourhood to \gaia\ data. According to \citet{2019MNRAS.490.1026H}, this model has a bar strength of ${\alpha_{\rm bar}=0.024}$ (for the definition of $\alpha_{\rm bar}$, see  Eq.~\ref{eq:alpha_bar_strength} in the following section). The bar might also have higher-order Fourier modes, such as ${\alpha_{\rm bar} (m=4) < 0}$ (see, e.g., \citealt{2018MNRAS.477.3945H}), which gives the bar a boxy appearance. But as these modes and their associated resonances are much weaker than the ${m=2}$ mode and essentially unconstrained, discussing them is beyond the scope of this paper.

\subsection{Running the barred test particle simulation}

In order to create our bar models, we evolve the initial axisymmetric model in a net potential (${\Phi_{\rm net} = \Phi_{\rm axi}+ \Phi_{\rm bar}}$), set by both the axisymmetric and a bar potential. We use the simple quadrupole (${m=2}$) bar model \citep{Dehnen2000,Monarietal2016}, which has a potential (${\Phi_{\rm bar}}$) of the following form
\begin{equation}
\begin{split}
\Phi_{\rm bar} (R, \phi, z)= A_{\rm b} (t) \cos(2(\phi - \Omega_{\rm bar}t)) \\
 \times \left(\frac{R}{r}\right)^2 \times
 \begin{cases}
 \left(\frac{-R_{\rm bar}}{r} \right)^3,& \text{if } r \geq R_{\rm bar}\\
 \left(\frac{r}{R_{\rm bar}} \right)^3 -2 , & \text{if } r \leq R_{\rm bar}\\
 \end{cases}
 \ ,
\end{split}
\label{eq:bar_potential_Dehnen}
\end{equation}
\noindent where $R_{\rm bar}$ is the extent of the bar, and $\Omega_{\rm bar}$ is the pattern speed of the bar. Later in this work, we will also refer to the bar orbital period, given by $T_{\rm bar} = 2\pi/ \Omega_{\rm bar}$. $A_{\rm b} (t)$ denotes the bar amplitude at time $t$, and has the following functional form:
\begin{equation}
A_{\rm b} (t) =
  A_{\rm f} \left( \frac{3}{16} \xi^5 - \frac{5}{8} \xi^3 +\frac{15}{16} \xi + \frac{1}{2} \right)
\,;\hspace{0.5 cm}
\xi \equiv 2 \frac{t}{t_1} - 1 \,.
\end{equation}
\noindent In this manner, the bar amplitude increases during the time-interval $ 0 < t \leq t_1$, and afterwards remains constant at $A_{\rm b} = A_{\rm f}$ \citep[for details, see][]{Dehnen2000}.

The bar strength ($\alpha_{\rm bar}$) is quantified via 
\begin{equation}
\alpha_{\rm bar} =  \frac{3 A_{\rm f}}{v_{\rm circ}^2 (R = R_{\odot})} \left (\frac{R_{\rm bar}}{R_{\odot}} \right)^3\,,
\label{eq:alpha_bar_strength}
\end{equation}
\noindent which describes the radial force ratio of the bar potential to the axisymmetric background potential, evaluated at the Solar radius, $R_{\odot}$, along the bar's major axis ($\phi = 45 \degrees$). The tracer particles have no mass and the stellar disc therefore has no self-gravity.

\subsection{The suite of barred simulations}
\label{sec:suite_of_bar_models}

Throughout this work, we vary $\alpha_{\rm bar}$ to create different bar models of different strengths. All these models use a bar pattern speed of ${\Omega_{\rm bar} = 40 \kmsk}$, which is similar to recent measurements of the bar pattern speed in the MW (see references in Section~\ref{sec:mw_bar}).
 We keep the length of the bar ($R_{\rm bar}$) fixed to ${4.5 \kpc}$ for all models, and the bar is always oriented along ${\phi = 45 \degrees}$. For the assumed potential, this pattern speed sets the corotation (CR) at ${R_{\rm CR} = 5.8 \kpc}$, and the Outer Lindblad Resonance (OLR) at ${R_{\rm OLR} = 8.9 \kpc}$. This, in turn, sets the parameter ${\mathcal{R} \equiv R_{\rm CR} / R_{\rm bar}}$ to 1.29, thereby qualifying the bar, present in our model, as a \textit{fast} ($\mathcal{R} < 1.4 $) bar \citep[for details, see e.g.,][]{DebattistaandSellwood2000}.
\par
In Sections~\ref{sec:pattern_speed_variation} and \ref{sec:survey_volume_effect} only, we will also consider a bar model where we set the pattern speed to ${\Omega_{\rm bar} = 28 \kmsk}$, and the strength of the bar to ${\alpha_{\rm bar}= 0.03}$. For the assumed potential, a pattern speed $\Omega_{\rm bar} = 28 \kmsk$ places the CR at ${R_{\rm CR} = 7.8 \kpc}$, and the OLR at ${R_{\rm OLR} = 12.1 \kpc}$. This, in turn, sets the parameter $\mathcal{R}$ to 1.7, thereby qualifying the bar, present in our model, as a \textit{slow}  ($\mathcal{R} > 1.4 $) bar.
\par
For all the models used here, we set $t_1 = 5 \, T_{\rm bar}$, and evolve the model for a total time of $50 \, T_{\rm bar}$. We used a symplectic integrator of order 6 for the orbit integration with time-step ($\Delta t$) equal to $T_{\rm bar} / 100$. To check the numerical accuracy of our integration scheme, we evolve the initial axisymmetric model over the same duration ($50 \, T_{\rm bar}$) under the influence of the axisymmetric potential ($\Phi_{\rm axi}$), finding that all the action components are \textit{conserved} (as expected) to within 1 percent.
\par
 Each model is referred as a unique string given by `{\sc [pattern speed][Fourier mode for bar][bar strength]}' where {\sc [bar pattern speed]} denotes the bar pattern speed (in $\kmsk$), {\sc [Fourier mode for bar]} denotes the Fourier coefficient used to model the bar ($m =2$ here for all models) while {\sc [bar strength]} denotes the bar strength ($\alpha_{\rm bar}$).  For example, \textbf{40m2a03} denotes a bar model with bar pattern speed of $40 \kmsk$, and bar strength $\alpha_{\rm bar} = 0.03$, while an $m=2$ Fourier mode is used in the bar potential. Fig.~\ref{fig:densmap_initialandfinal} (right-hand panels) shows the corresponding density distribution of stars at $t = 50 \ T_{\rm bar}$ in a face-on projection (i.e., in the ${(x,y)}$-plane) as well as in an action-space projection (the ${(J_R, L_z)}$-plane) for the model \textbf{40m2a02}.

\subsection{The suite of mock datasets from different survey volumes}
\label{sec:survey_volume}

As motivated in section~\ref{sec:Intro}, we will apply \roadmapping\ to mock datasets drawn from the barred models described in section~\ref{sec:bar_model}, in order to investigate how the dynamical influence  of the stellar bars on discs of MW-like galaxies might or might not affect the robustness of action-based dynamical modelling (see section~\ref{sec:results}). 
\par
For this work, we use the test particle simulations described in section~\ref{sec:suite_of_bar_models}, and always take a spherical survey volume centred on a hypothetical observer at ${\bf{x_{\rm cen}} = (R_{\rm cen}, \phi_{\rm cen}, z_{\rm cen})}$,  with radius ${d_{\rm max} = 4 \kpc}$. This mimics the observations of bright stellar standard candles in the absence of dust extinction by a contiguous survey like \gaia, and uses the 6-D phase-space information of the simulated stars falling in that selected survey volume as the input data for \roadmapping.
\par
In the final snapshot of our models, the bar major axis is always oriented along ${\phi = 45 \degrees}$, and the minor axis along ${\phi = - 45 \degrees}$. For each bar model, we select three survey volumes, one each along the bar major and minor axes, and third one falling between them (at ${\phi = 0 \degrees}$). The latter can be considered as being closest to our Sun's location with respect to the Galactic bar, which is roughly ${\sim \! 28 \! - \! 33 \degrees}$ with respect to the Solar neighbourhood \citep[e.g.,][]{Weggetal2015}. For all three regions, we set ${R_{\rm cen} = 8.2 \kpc}$ and ${z_{\rm cen} = 0}$. We use the suffixes `\textbf{Major}', `\textbf{Minor}', and `\textbf{Inter}' after a bar model's name to denote whether the given survey volume falls along the bar major, minor axis, or between them, respectively. 
\par
For each selected survey volume, we randomly select 7,250 stars for our \roadmapping\ analysis. We point out that within a survey volume $\mathcal{D}$ with radius ${d_{\rm max} = 4 \kpc}$, there are ${\sim \!\! 33,000}$ stars. However, in this work, we also study the effect of the survey volume's coverage on the accuracy of the recovered best-fit parameters, by shrinking $d_{\rm max}$ to 3 and ${2 \kpc}$ (for details, see section~\ref{sec:survey_volume_effect}). In order to achieve a fair comparison of the best-fit parameters and their associated errors, the total number of (randomly) selected stars is kept fixed for all cases explored here. 
\par
In our bar models, we occasionally encounter a small number (less than 1\%) of stars entering in our selected survey volume that are on extreme orbits (e.g., moving on nearly perfectly radial trajectories toward the galactic centre or counter-rotating with significant angular momentum). These outliers with very low likelihoods can potentially impact the \roadmapping\ modelling. To limit the influence of these stars, we manually set a floor value on the stellar likelihoods of $ \mathcal{L}_{\rm floor} = 10^{-4} \times {\rm median} (\mathcal{L}_i)$, where the median is taken over all stars in the survey volume. Noe that this floor changes from step to step, as it depends on the distribution of stellar likelihoods. A similar approach was also taken in \citet[see section 3.1 there]{Tricketal2017}.

\section{RoadMapping modelling \& effect of bar resonances in action space}
\label{sec:roadmapping}
%
Here, we briefly describe \roadmapping, which is an action-based dynamical modelling framework to simultaneously recover the parameters of the underlying DF as well as the (axisymmetric) potential from discrete stellar tracers in the MW disc. Additionally, we provide some background on the effect of bar resonances and how they can possibly modify the distribution of stars in action space.

\roadmapping\ is an acronym that stands for ``Recovery of the Orbit Action Distribution of Mono-Abundance Populations and Potential INference for our Galaxy''.
This machinery was inspired by the findings of \citet{Bovyetal2012}, that the MW's disc can be considered as a superposition of several stellar mono-abundance populations (MAPs) with roughly exponential profiles, and of \citet{Tingetal2013}, who showed that MAPs can be modelled with the action-based qDF by \citet{BinneyandMcMillan2011}. \citet{BovyandRix2013} applied a first version of \roadmapping\ to MAPs from SEGUE \citep{Yannyetal2009}, providing one of the best measurements of the MW's surface density profile to date. The \roadmapping\ machinery was subsequently rewritten, extended, and thoroughly tested by \citet{Tricketal2016}. For further details, we refer the reader to \citet{BovyandRix2013} and \citet{Tricketal2016}.

\subsection{Summary of the algorithm}

For an assumed axisymmetric potential, $\Phi_\text{axi}(R, z)$, with a set of parameters $p_{\Phi}$, the probability density that the \textit{i}th star is associated with an orbit having the actions ${\bf{J}}_i$ is proportional to the assumed distribution function qDF(${\bf{J}}$), itself dependent on a set of parameters $p_{\rm DF}$. Therefore, the joint likelihood of a star being within a survey volume $\mathcal{D}$, and on a given orbit is (following \citealt{Tricketal2016,Tricketal2017}) given by
\begin{equation}
\mathcal{L}_i = \frac{ {\rm qDF} ({\bf x}_i, {\bf v}_i | p_{\Phi}, p_{\rm DF})}  { \int_{\mathcal{D}} {\rm qDF} ({\bf x}, {\bf v} | p_{\Phi}, p_{\rm DF}) \times {\rm SF} ({\bf x}) \ d^3 x \ d^3 v}\,.
\label{eq:roadmapping}
\end{equation}
Throughout this work, we use a purely spatial and volume-complete data selection function, SF$(\bf{x})$, defined as 
\begin{equation}
\rm {SF} (\bf{x}) \equiv
\begin{cases}
1, \rm{if} \ \bf{x} \in \mathcal{D}\\
0, \rm {otherwise}
\end{cases},
\label{eq:selection_func}
\end{equation}
\noindent where $\mathcal{D}$ is the selected spherical survey volume (for details, see section~\ref{sec:survey_volume}). The denominator in Eq.~\ref{eq:roadmapping} is equal to the expected total number of stars enclosed in that survey volume $\mathcal{D}$, given the model parameters $p_{\Phi}$ and $p_{\rm DF}$. This integral is evaluated (with high numerical accuracy) using the technique described in section~\ref{sec:generate_equi_model}. In addition, following \citet{Tricketal2016,Tricketal2017}, we assume flat priors on the model parameters $p_{\Phi}$ and $\ln p_{\rm DF}$.
Then, using the Bayesian approach, we find the approximate location of the maximum and the width of the posterior probability density function $p (p_{\Phi}, p_{\rm DF} \, | \, {\rm data}) \propto {\rm prior} (p_{\Phi}, p_{\rm DF}) \, \prod_{i=1} ^{N} \mathcal{L}_i$ using a nested-grid formalism that accounts for the fact that changing the potential parameters is computationally much more expensive than changing the DF parameters.\footnote{This is because changing the potential requires the mapping between actions and phase-space coordinates to be recomputed.} In a second step, we initialize a Monte Carlo Markov Chain (MCMC) sampler\footnote{For this work, we make use of the MCMC software {\sc emcee} by \citet{Foreman-Mackeyetal2013}} at the peak of the posterior density, which then explores and recovers the full shape of the posterior distribution. For more details, we refer the reader to \citet{Tricketal2016}. For each \roadmapping\ analysis, we set the number of grid refinements ($N_{\rm ite}$) in the initial grid search to 20, the number of MCMC burn-in steps ($N_{\rm burnin}$) to 200 per walker, and the number of subsequent MCMC steps ($N_{\rm MCMC}$) to 500 per walker \citep[for details, see][]{Tricketal2016}. We use the affine-invariant sampler of \citet{GoodmanWeare2010} with $64$ walkers.
  
\subsection{Background on the bar resonances in action space} 
\label{sec:recap_resonances_action_space}

The effect of a bar on orbits in axisymmetric action space has been studied extensively in the past (e.g. \citealt{SellwoodBinney2002,2018MNRAS.474.2706B,2020MNRAS.495..886B,2019MNRAS.490.1026H,2019A&A...626A..41M,Tricketal2021}). Here we summarize the essentials, as this helps to understand what drives systematic biases in \roadmapping.
\par
A star orbiting in the midplane of an axisymmetric potential has two fundamental frequencies: a radial frequency $\Omega_{R}$, that governs the star's oscillation between pericenter and apocenter, and an azimuthal frequency $\Omega_{\phi}$, that describes the azimuthal part of the orbital motion around the Galactic centre. For a generic potential $\Phi_\text{net} = \Phi_\text{axi} + \Phi_\text{bar}$, these frequencies no longer fully describe the orbit, and can only be determined via a Fourier Transform of the integrated orbit. In a generic, non-axisymmetric potential, we denote these frequencies $\Omega_{R,\text{true}}$ and $\Omega_{\phi,\text{true}}$. The bar itself rotates with a frequency $\Omega_\text{bar}$. If these three frequencies are integer multiples of each other, i.e. if the orbit is in \textit{resonance} with the bar, 
\begin{equation}
    m \cdot (\Omega_{\rm bar} - \Omega_{\phi, \rm true}) - l\cdot \Omega_{R, \rm true} =0\, \text{(resonance condition)}\,,\label{eq:true_resonance_condition}
\end{equation}
and the bar has a similar effect on the star's orbit as walking on a bridge with the bridge's eigenfrequency would have. The bar can induce additional oscillations in the orbit and in action space that have a large amplitude, the so-called \textit{libration}. The ILR corresponds to $l = -1, \ m = 2$, whereas the OLR corresponds to  $l = 1, \ m = 2$, and the CR corresponds to $l=0$.

Even if we have imperfect knowledge of the true potential, and therefore imperfect knowledge of the true stellar orbits, we can still estimate the location of the bar resonances in action space using the axisymmetric frequency estimates $\Omega_{\phi,\text{axi}}$ and $\Omega_{R,\text{axi}}$, i.e. the orbital frequencies the star \textit{would} have in the axisymmetric background potential $\Phi_\text{axi}$ given its current $(\vec{x},\vec{v})$. These axisymmetric frequency estimates are functions of the actions alone, and only correspond to the true orbital azimuthal and radial frequencies in the absence of a bar. The condition
\begin{equation}
m \cdot (\Omega_{\rm bar} - \Omega_{\phi, \rm axi}) - l\cdot \Omega_{R, \rm axi}
  = 0 \ \text{(axisymmetric resonance line)}
\label{eq:resonance_line}
\end{equation}
(for ${J_z=0}$) is satisfied along lines in action space, the \textit{axisymmetric resonance lines} (hereafter ARLs). For a pure quadrupole bar, we expect the $m=2$ resonances to be dominant. In Fig.~\ref{fig:action_space_resonances_weak_strong} (and also Fig.~\ref{fig:densmap_initialandfinal}), we therefore overplot the resonance lines (ARL) for $m=2$ and ${l\in \{0,1,2,3\}}$. Resonant orbits librate back and forth in action space, always centred on the resonance lines (see e.g. \citealt{2018MNRAS.474.2706B,2020MNRAS.495..886B,Tricketal2021}). In Fig. \ref{fig:action_space_resonances_weak_strong}, we use contours to mark the location of resonant stars in the test particle simulation, where the true orbit frequencies satisfy Eq.~\eqref{eq:true_resonance_condition} (and an additional cut in $\Omega_{R,\text{true}} > 15 \kmsk$ to exclude near-circular orbits). This indicates the size of the libration region around the resonance line. As shown by \citet{SellwoodBinney2002}, the libration amplitudes $\delta L_z$ and $\delta J_R$ are coupled according to the relation
\begin{equation}
    \frac{\delta J_R}{\delta L_z} = \frac{l}{m}.
\end{equation}
\par
This relation nicely illustrates how the bar resonances redistribute stars in action space.
At CR ($l=0$), $\delta J_R/\delta L_z = 0$ but $\delta L_z$ can still become quite large. This means that no substantial orbit \textit{heating} (related to an increase in $J_R$) is occurring, but that a great deal of orbital \textit{radial migration} occurs across the CR, both towards smaller and larger $L_z$. This \textit{churning} leads to flatter density profiles in both $L_z$ and radius $R$ in the stellar disc. At the OLR ($l=1, m=2$), a change in $L_z$ is coupled to a change in $J_R$ by $\delta J_R/\delta L_z = 1/2$. As in an exponential stellar disc more stars live at small $L_z$ and small $J_R$ than at large $L_z$ and large $J_R$, we get a net redistribution of stars from the lower left across the OLR resonance line to the upper right. This creates the distinct high-$J_R$ ridge to the right of the OLR resonance line that can be seen in Fig.~\ref{fig:action_space_resonances_weak_strong} (see also Fig.~\ref{fig:densmap_initialandfinal}). The stronger the bar, the larger the maximum libration amplitude $\delta L_z$ and $\delta J_R$ that orbits can have at a resonance, the stronger the resonant ridges and the larger the regions in action space that are dominated by resonant orbits that \roadmapping\ intrinsically cannot model.

\section{Results}
\label{sec:results}

\subsection{The influence of bar strength}
\subsubsection{No bar} 

Before we evaluate the dynamical influence of bars on the recovery of the parameters of the (axisymmetric) potential and the DF, we perform a simple sanity test of \roadmapping's performance when applied to an axisymmetric model evolved for a same total time as any of the bar models ($50 \ T_{\rm bar}$). We choose a survey volume along $\phi  = 0$ (with $d_{\rm max} = 4 \kpc$, and $R_{\rm cen} = 8.2 \kpc$) from the model \textbf{40m2a00}, and apply \roadmapping\ on it. The resulting best-fit values of the parameters are quoted in Table~\ref{tab:bestfit_params_axisymmetric}. As seen clearly from Table~\ref{tab:bestfit_params_axisymmetric}, the precision for $7,250$ stars is better than $1$ percent of their corresponding true values. The parameters are recovered within ${1 \, \sigma}$. This demonstrates that \roadmapping\ can accurately recover the potential and the DF parameters when it is applied to an axisymmetric model which has been evolved for a certain time-span under the (same) axisymmetric potential.

\begin{table}
\caption{Best-fit parameters for \textbf{40m2a00} survey volume. This illustrates the accuracy and precision of the \roadmapping\ fit in absence of bar resonances in the data.}
\begin{tabular}{ccc}
\hline
\hline
\multicolumn{3}{c}{\textbf{Potential parameters ($p_{\Phi}$)}} \\
\hline
 $v_{\rm circ}(R = R_{\odot})$ (\kms) & : & $220.03 \pm 0.31$\\
$a_{\rm disc}$ (\kpc) & : & $2.47  \pm  0.04$\\
$b_{\rm disc}$ (\kpc) & : & $0.3 \pm 0.02$\\
\hline
\multicolumn{3}{c}{\textbf{DF parameters ($p_{\rm DF}$)}} \\
\hline
$\ln(h_{R}/8 \kpc )$ & : & $-1.16  \pm  0.02$\\
$\ln(\sigma_{R,0}/220 \kms )$ & : & $-1.89  \pm  0.01$\\
$\ln(\sigma_{z,0}/220 \kms)$ & : & $ -2.18  \pm  0.02$\\
\hline
\hline
\end{tabular}
\label{tab:bestfit_params_axisymmetric}
\end{table}
\subsubsection{A weak bar} 
\label{sec:weak_bar}

\begin{figure*}
\includegraphics[width=1.1\linewidth]{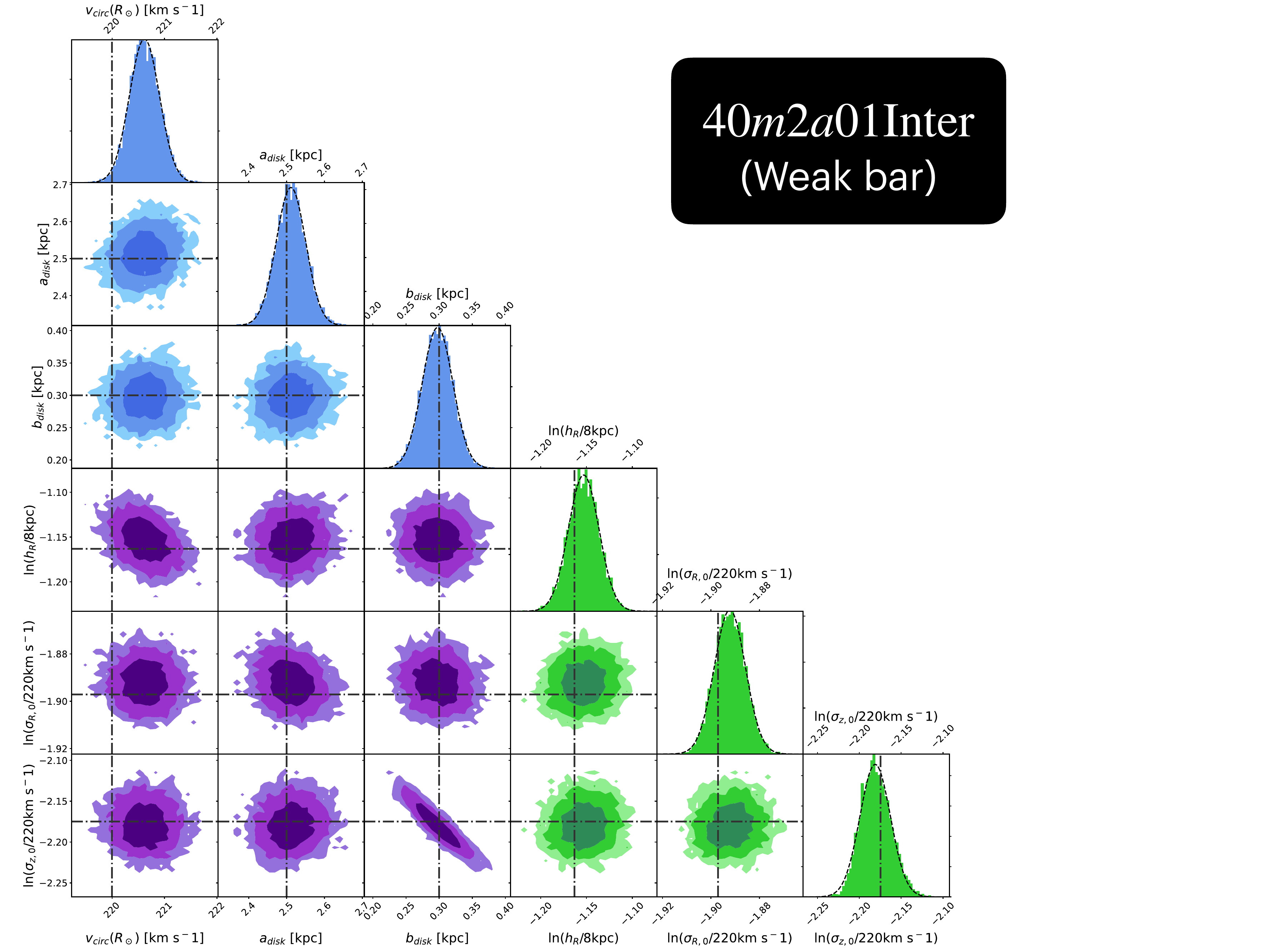}
\caption{Posterior densities of the best-fit parameters $p_{\rm M} \equiv \{ p_{\Phi}, p_{\rm DF} \}$ recovered with \roadmapping\ from the \textbf{40m2a01Inter} dataset. This dataset was selected from a region $45 \degrees$ behind a weak bar. Blue coloured histograms indicate the posteriors for the potential parameters ($p_{\Phi}$), green coloured histograms indicate the qDF parameters ($p_{\rm DF})$, and purple histograms indicate joint distributions involving both potential and qDF parameters. The true values of the parameters are indicated by the horizontal/vertical dash-dotted lines. The dark, medium, and bright contours in the 2D distributions represent $1 \sigma$, $2 \sigma$, and $3 \sigma$ confidence regions, respectively. We use $N_{\rm ite} = 20$, $N_{\rm MCMC} = 500$, and $N_{\rm burnin} = 200$ (for details, see section~\ref{sec:roadmapping}). Even in presence of a weak bar,  the potential and the DF parameters, are recovered fairly accurately (within $1-3 \sigma$).
}
\label{fig:cornerplot_m2a01reg03}
\end{figure*}
\begin{figure}
\centering
\subfigure[Weak bar.]{\label{fig:reso_overlap_weak}\includegraphics[width=0.8\columnwidth]{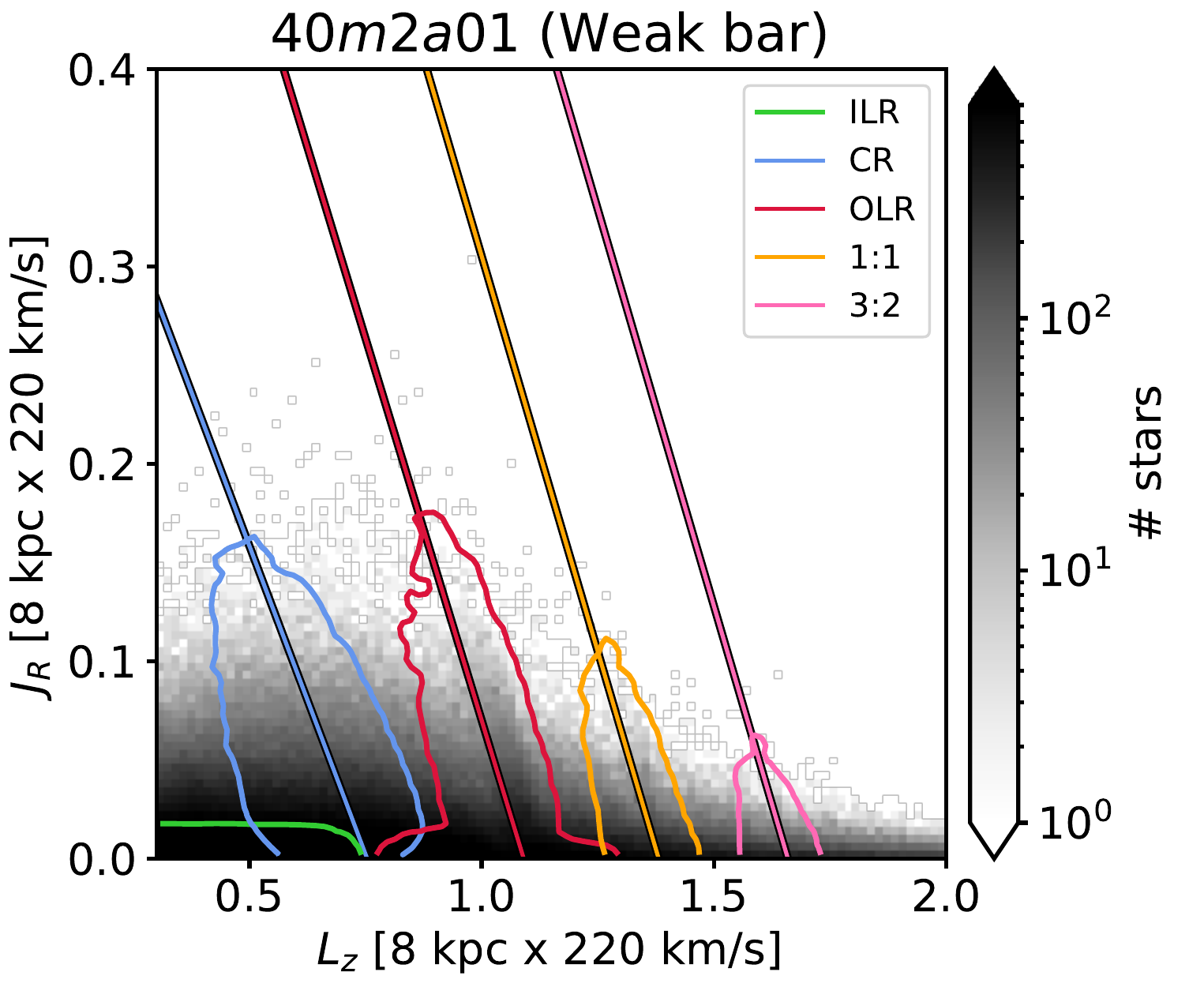}}
\subfigure[Strong bar with resonance overlap.]{\label{fig:reso_overlap_strong}\includegraphics[width=0.8\columnwidth]{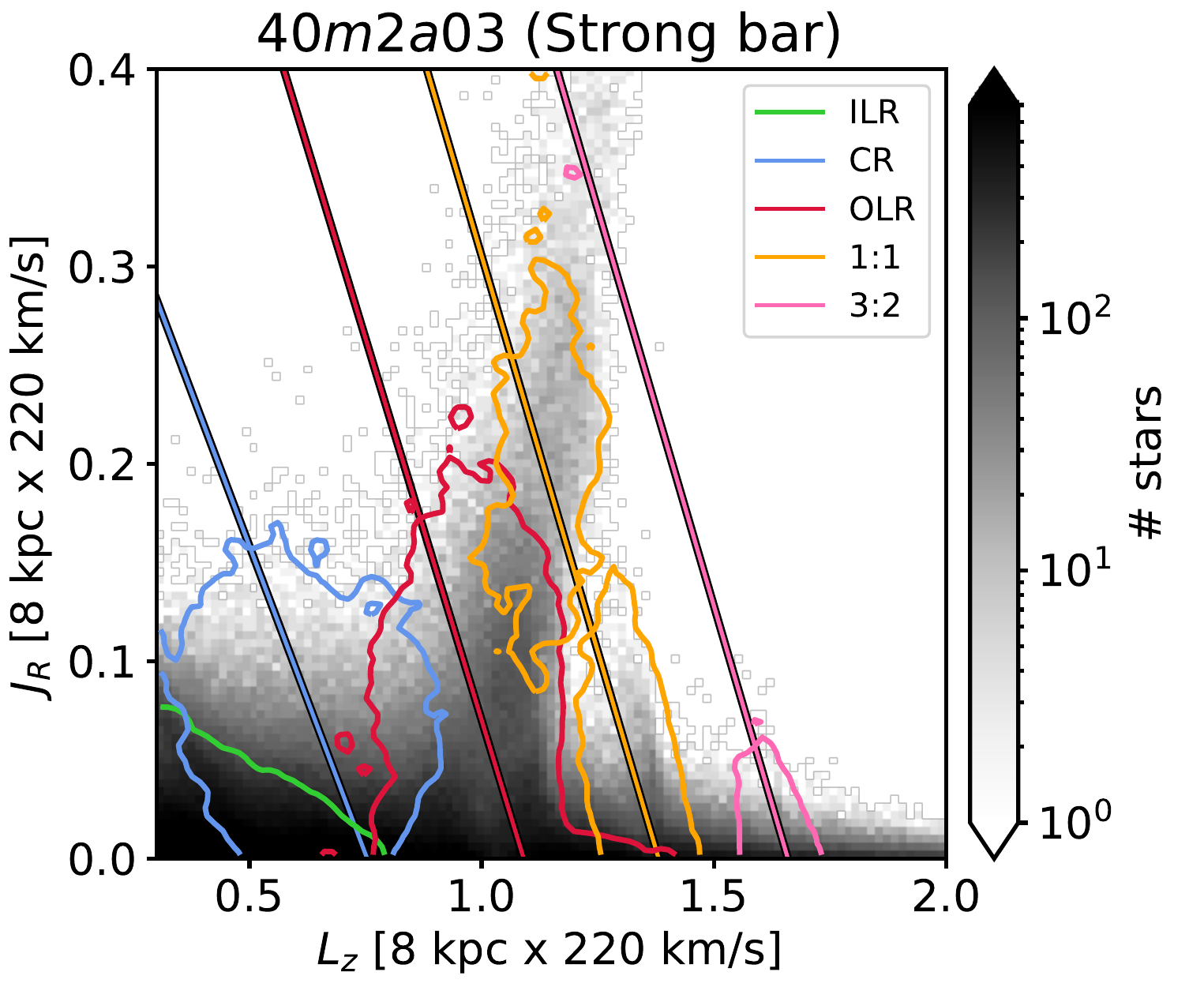}}
\caption{Distribution of the stars in the ${(J_R, L_z)}$-plane for the barred models \textbf{40m2a01} (top panel) and \textbf{40m2a03} (bottom panel), for which the corresponding \roadmapping\ results are shown in Figs.~\ref{fig:cornerplot_m2a01reg03} and \ref{fig:cornerplot_m2a03reg03}, respectively. The negatively sloped lines show the \textbf{ARLs} from Eq.~\eqref{eq:resonance_line}, and the (smoothed) contours indicate where the resonant stars live according to Eq.~\eqref{eq:true_resonance_condition}, i.e. the libration regions. For a bar as strong as ${\alpha_{\rm bar}=0.03}$ with pattern speed ${40 \kmsk}$, the libration amplitudes are so large that the resonances begin to overlap, which leads to the growth of an extremely dominant ridge next to the OLR resonance line. The ARL corresponding to the ILR lies off the left of the plot (in both panels), but stars trapped at ILR are indicated by green contours.
}
\label{fig:action_space_resonances_weak_strong}
\end{figure}

Next, we choose the weakest bar model, i.e., \textbf{40m2a01}. This model was also investigated in \citet{Tricketal2019}, who showed that even with this bar strength, the bar already causes visible resonance signatures in the disc. We then apply \roadmapping\ to the survey volume that falls in-between the bar major and minor axes, i.e., to the dataset \textbf{40m2a01Inter}. The resulting best-fit values of the parameters and their associated posterior distributions are shown in Fig.~\ref{fig:cornerplot_m2a01reg03}. Even a visual inspection reveals that the potential and the DF parameters are recovered reasonably well by \roadmapping. For a bar strength of ${\alpha_{\rm bar}=0.01}$, \roadmapping\ fully recovers the underlying global axisymmetric potential within $1-2 \sigma$, and is therefore robust against the presence of the bar's OLR and CR in the data. Fig.~\ref{fig:reso_overlap_weak} shows the distribution of stars in an action-space projection (specifically, the ${(J_R,L_z)}$-plane) as well as the positions of the corresponding ARLs. The distribution of resonant stars (with $m=2$ and $l\in \{0,1,2,3\}$) in the ${(J_R,L_z)}$-plane is also marked. As seen clearly, although the weak bar produces ridge-like features in the density distribution of stars in the ${(J_R,L_z)}$-plane, the distribution of resonant stars are still spatially well-separated/decoupled in the same action plane (see coloured contours in Fig.~\ref{fig:reso_overlap_weak}). In other words, the dynamical phenomenon of resonance overlap \citep[for details, see e.g.][]{BT08} does not occur for the weak bar model \textbf{40m2a01}.
\par
With one exception, there are no strong correlations between the recovered model parameters. The strong anti-correlation between $\sigma_{z,0}$ and $b_{\rm disc}$ is due to the following physical reason. Intuitively, a smaller $b_{\rm disc}$ value would set a steeper gradient in the potential along the vertical direction $\left( \frac{\partial \Phi}{\partial z} \right)$. Now, the Jeans equation, when applied to a system whose DF is of the form $f(\mathcal{H}, L_z)$, $\mathcal{H}$ being the Hamiltonian, shows that the vertical velocity dispersion is directly proportional to the potential gradient in the vertical direction \citep[for details, see][]{BT08}. This explains why a steeper potential gradient (with respect to $z$) always yields a higher $\sigma_{z,0}$ value, and thus making the quantities $\sigma_{z,0}$ and $b_{\rm disc}$ strongly anti-correlated.

\subsubsection{A strong bar} 
\label{sec:strong_bar}
\begin{figure*}
\includegraphics[width=1.1\linewidth]{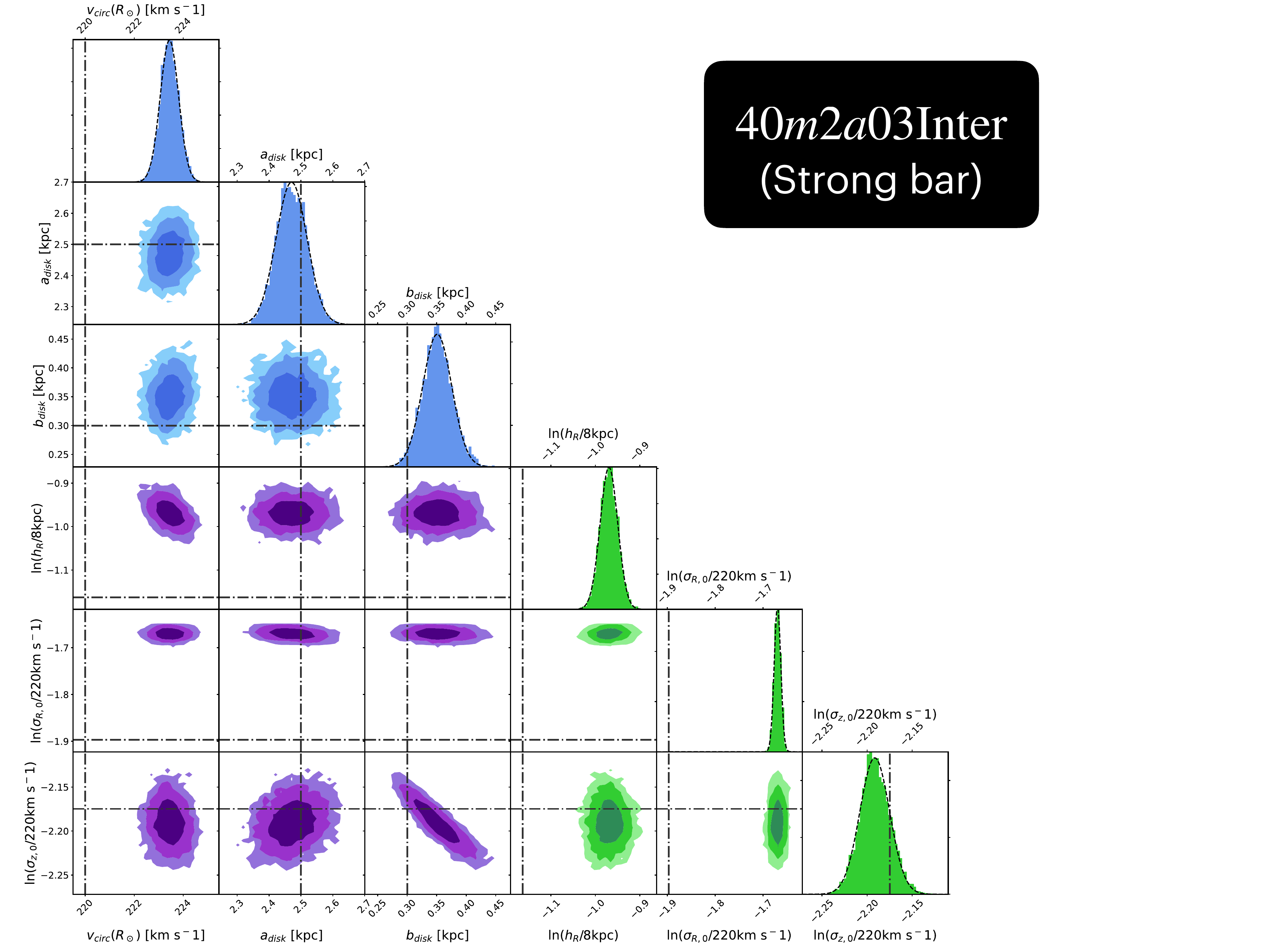}
\caption{Same as Fig.~\ref{fig:cornerplot_m2a01reg03}, but for the dataset \textbf{40m2a03Inter}, which was selected from a disc with a \textit{strong} bar (and which is located $45 \degrees$ behind the bar). In presence of a strong bar, two of the potential parameters, $v_{\rm circ}$ and $a_{\rm disc}$,  are still recovered within $\sim \! 1 - \! 2$ percent relative error, while the third potential parameter, $b_{\rm disc}$, is recovered within $\sim 17$ percent relative error.  The relative errors in recovering the DF parameters ($\ln (h_R), \ \ln(\sigma_{R,0})$) are fairly large as well (${\sim \! 12 \! - \! 17}$ percent). This is because in this model a strong bar induces resonance overlap, which in turn, facilitates the growth of strong ridge-like features in action space. The action-space distribution of the strong bar model thus deviates significantly from the axisymmetric model (for details, section.~\ref{sec:strong_bar}).
}
\label{fig:cornerplot_m2a03reg03}
\end{figure*}

Lastly, we examine the performance of \roadmapping\ for a strong bar model (\textbf{40m2a03}), using the same survey volume that falls between the major and minor axes, i.e.,  \textbf{40m2a03Inter}. The resulting best-fit values of the parameters and their associated posterior distributions are shown in Fig.~\ref{fig:cornerplot_m2a03reg03}. 
A mere visual inspection of Fig.~\ref{fig:cornerplot_m2a03reg03} reveals that there are deviations from the true values in the recovered potential and DF parameters. To express these discrepancies quantitatively, the precision (i.e. the absolute error = standard deviation derived from the pdf) for 7,250 stars is the same as for the no-bar case as shown in Table~\ref{tab:bestfit_params_axisymmetric}. Two of the potential parameters, $v_{\rm circ}$ and $a_{\rm disc}$,  are still recovered within $\sim 1-2$ percent relative error, while the third potential parameter, $b_{\rm disc}$, is recovered within $\sim 17$ percent relative error. The relative errors in the DF parameters describing the in-plane stellar orbit distributions are larger than those of the potential parameters. To illustrate this point, for the radial scale length ($h_R$) and the radial velocity dispersion at $R_{\odot}$ ($\sigma_{R,0}$), the relative errors are $\sim 12-17$ percent. In contrast, the vertical velocity dispersion at $R_{\odot}$ ($\sigma_{z,0}$) is recovered fairly accurately (with relative error of $\sim 1$ percent). In terms of the systematic biases in recovering the parameters, the potential parameters are recovered within ${1-9 \, \sigma}$, whereas the radial scale length ($h_R$) is recovered to within $9 \, \sigma$ and the radial velocity dispersion at $R_{\odot}$ ($\sigma_{R,0}$) is recovered to within $23 \, \sigma$. The vertical velocity dispersion at $R_{\odot}$ ($\sigma_{z,0}$) is recovered fairly well (to within $1 \, \sigma$).
In the strong bar case, the formal uncertainties reported by \roadmapping\ no longer correspond to the actual scale of the residuals (inferred minus true parameters). This is a consequence of model mismatch: as the non-axisymmetric features in the disc become stronger, the actions calculated with the axisymmetric potential become less relevant, and hence the axisymmetric model assumed by \roadmapping\ has a harder time capturing the underlying dynamics.
%
%
\par
To investigate why a stronger bar introduces more formal uncertainties for our best-fit axisymmetric model, we compute the density distribution of stars in the ${(J_R,L_z)}$-plane, and also determine the distribution of stars in resonance with $m=2$ and ${l \in \{0,1,2,3\}}$. This is shown in Fig.~\ref{fig:reso_overlap_strong}. As seen from Fig.~\ref{fig:reso_overlap_strong}, the model develops a strong ridge in the ${(J_R,L_z)}$-plane (for a physical explanation, see section~\ref{sec:recap_resonances_action_space}). In addition, the stars in resonance with the CR overlap with the stars that are in resonance with 1:2 OLR. A similar resonance-overlap phenomenon is also seen between the 1:2 OLR and the 1:1 OLR. This resonance overlap facilitates the substantial growth of the ridge (in the action plane) next to the 1:2 OLR resonance line (see Fig.~\ref{fig:reso_overlap_strong}). For further details, see Appendix~\ref{appen_ridgeFormation}.

\subsubsection{Recovering the distribution function}
\label{sec:recover_df}
Here, we briefly discuss the physical effect of the non-axisymmetric bar feature on the recovery of the DF parameters via \roadmapping. As demonstrated in the previous section, with increasing bar strength, the recovery of DF parameters suffers from increasing systematic deviations from their true values. Here, we study the underlying dynamical cause for this phenomenon.
\par
\begin{figure*}
\textbf{\Large {40m2a01Inter (Weak bar)}}
\includegraphics[width=\linewidth]{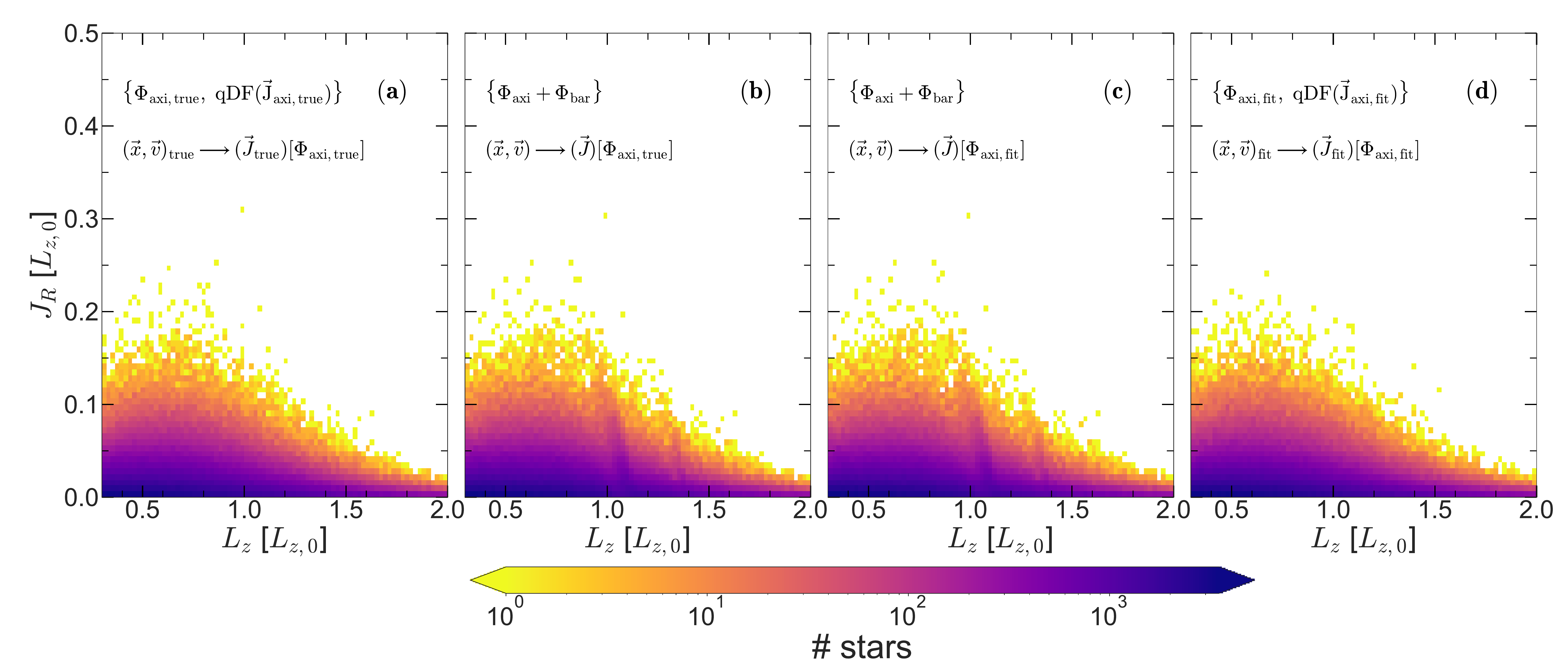}
\vspace{0.15cm}\\
\textbf{\Large {40m2a03Inter (Strong bar)}}
\includegraphics[width=\linewidth]{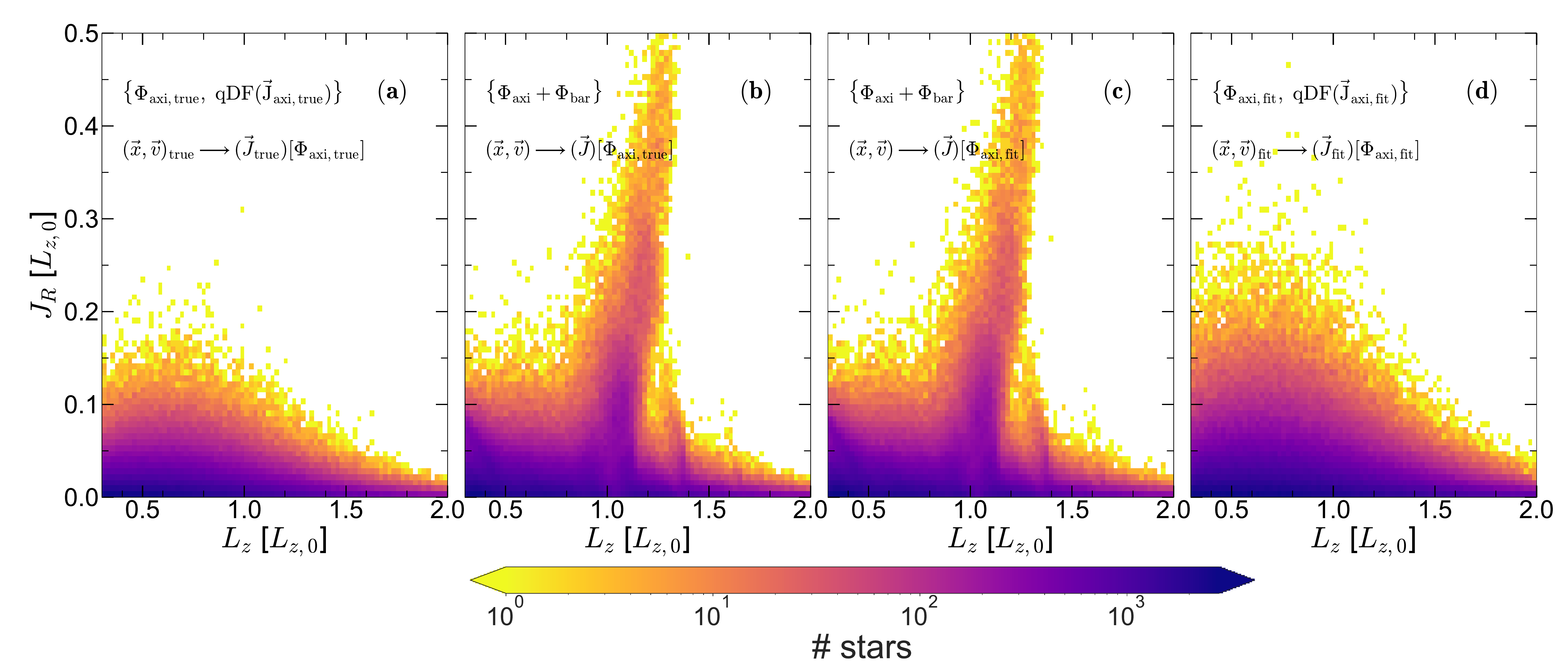}
\caption{Distribution of stars in action space: panel \textbf{(a)} shows all disc stars generated from the initially axisymmetric potential and smooth qDF. The actions are calculated using the true axisymmetric potential. This distribution is the `axisymmetric ground-truth' that we wish to recover with \roadmapping. Panel \textbf{(b)} shows the disc stars in the test particle simulation after the orbits have been integrated in the combined potential of the axisymmetric model plus the bar potential for ${\sim \! 8.4 \Gyr}$. The actions are still calculated using the true axisymmetric background potential. These $(x,v)$ values enter \roadmapping\ as input data. Panel \textbf{(c)} uses the same $(x,v)$ values as panel (b), but the actions are now calculated using the best-fit axisymmetric potential from \roadmapping. Finally, panel \textbf{(d)} shows an axisymmetric distribution of stars generated using \roadmapping{}'s best-fit axisymmetric potential and the  best-fit qDF. The actions are also calculated using the best-fit axisymmetric potential. During the fit, \roadmapping\ attempts to minimize the difference between panels \textbf{(c)} and \textbf{(d)}, by varying the qDF and $\Phi_{\rm axi}$ parameters. The top row illustrates the results of applying \roadmapping\ to the \textbf{40m2a01Inter} dataset, while the bottom row shows the results for the \textbf{40m2a03Inter} dataset.
}
\label{fig:effect_on_actions}
\end{figure*}

As summarized in section~\ref{sec:recap_resonances_action_space} and illustrated in Fig.~\ref{fig:action_space_resonances_weak_strong}, a strong bar creates strong high-$J_R$ ridges at the OLR, and induces a strong redistribution of stars in $L_z$ at CR. 
\roadmapping\ makes use of the qDF, which is a function of the actions (see Eq.~\ref{eq:qDF}). In Fig.~\ref{fig:effect_on_actions}, we sequentially demonstrate how the stellar distribution in action space is computed at different stages of the \roadmapping\ analysis. After evolving the initial axisymmetric model for a certain time under the influence of the bar potential, ridges appear in action space (compare panels (a) and (b) in Fig.~\ref{fig:effect_on_actions}). The qDF, which is used in the \roadmapping\ framework, is unable to accurately capture these ridges. Hence, \roadmapping\ accommodates these ridges by expanding the entire DF towards larger $J_R$ values (compare panels (a) and (d) in Fig.~\ref{fig:effect_on_actions}). This leads to a large deviation (from the value corresponding to the axisymmetric configuration) of the radial velocity dispersion, i.e. the DF parameter $\ln \sigma_{R,0}$. Due to the flattening of the stellar density profile at CR as well as through the net-shift of stars towards larger $L_z$ at the OLR, the radial tracer scale length $h_R$ thus deviates from the value corresponding to the axisymmetric configuration when fitting the qDF.
For the weak bar, these ridges and redistributions are weaker than those for the strong bar model (compare top and bottom panels of Fig.~\ref{fig:effect_on_actions})). In other words, the underlying axisymmetric configuration is not dramatically altered due to the dynamical effect of a weak bar. Therefore, the DF parameters are better recovered/constrained for the weak bar model (as compared to the strong bar model).

\subsubsection{Recovering the potential}
\label{sec:recovery_potparams}

\begin{figure*}
\includegraphics[width=0.95\linewidth]{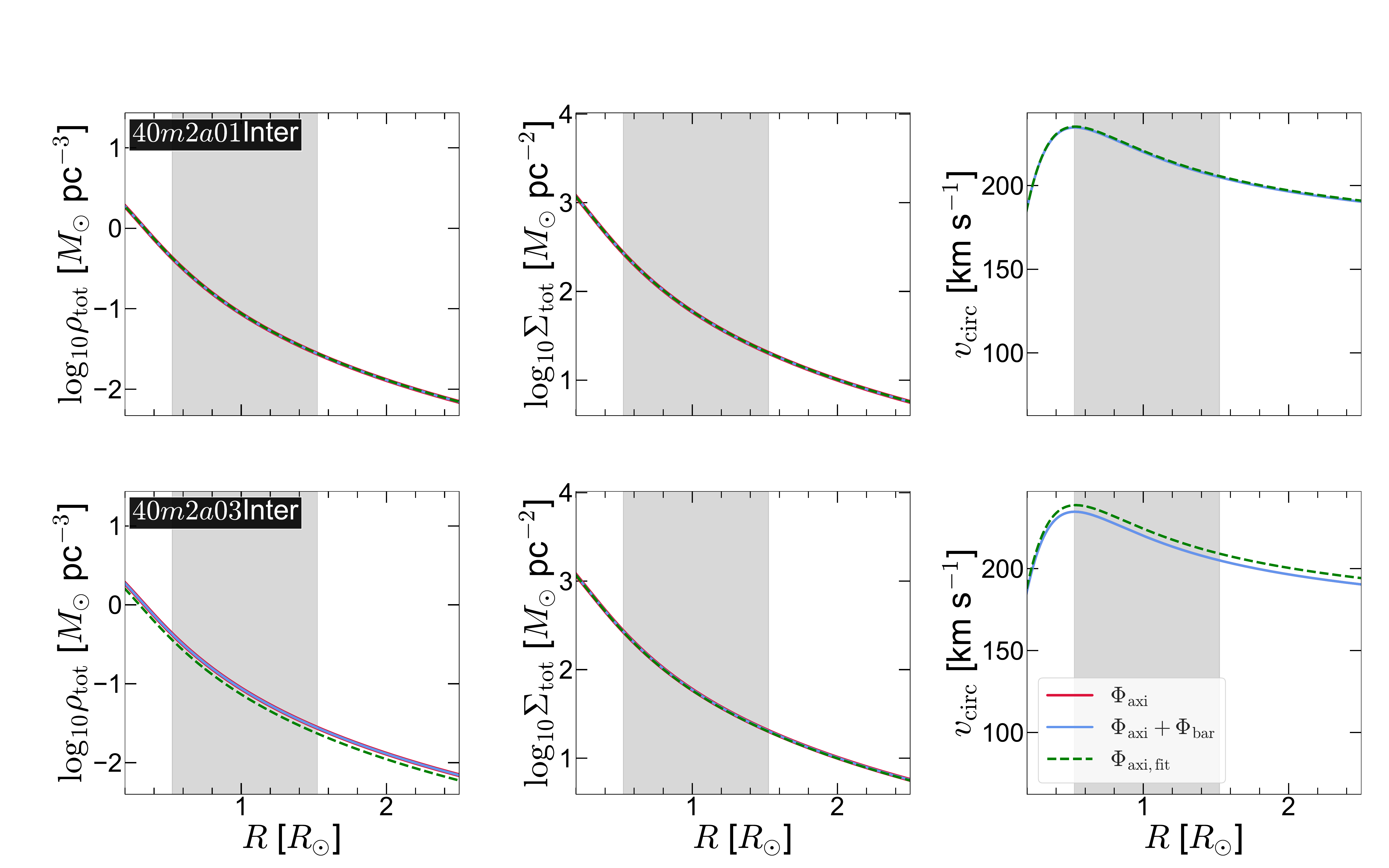}
\caption{Recovery of potential parameters. Here, we compare the true $\Phi_{\rm axi}$ with the true ${\Phi_{\rm axi} + \Phi_{\rm bar}}$ (evaluated at ${\phi = 0 \degrees}$) and the best-fit $\Phi_{\rm axi, fit}$ potential. \textit{Left panels}: radial distributions of the total mid-plane volume density, $\rho(R, z=0)$, related to the potential via the Poisson equation. \textit{Middle panels}: radial distributions of the surface density, obtained by integrating $\rho (R,z)$ within ${z = \pm 1 \kpc}$. This quantity is related to the vertical force ${F_z=\partial \Phi / \partial z}$.
\textit{Right panels:} radial distributions of the circular velocity ($v_{\rm circ}$), related to the radial force ${F_R = \partial \Phi / \partial R}$. The \textit{top row} shows the case of a weak bar model (\textbf{40m2a01Inter}), whereas the \textit{bottom row} shows the case of a strong bar model (\textbf{40m2a03Inter}). The vertical shaded region (in grey) denotes the radial extent of the survey volume. The forces (equivalent to the first derivatives of the potential) are globally recovered reasonably well (within ${1 \! - \! 2}$ percent relative error), even in presence of a strong bar (see bottom row). As throughout this work, ${R_{\odot} = 8 \kpc}$.
 }
\label{fig:residualmaps}
\end{figure*}

Here, we briefly discuss the physical effect of the non-axisymmetric bar resonances on the recovery of the potential via \roadmapping. In Fig.~\ref{fig:residualmaps}, we show the radial variation of the mid-plane volume density, surface density, and circular velocity ($v_{\rm circ}$), calculated for the true axisymmetric potential ($\Phi_{\rm axi}$), true axisymmetric+bar potential ($\Phi_{\rm axi} + \Phi_{\rm bar}$), and the best-fit axisymmetric potential ($\Phi_{\rm axi, fit}$) as obtained using \roadmapping. For the non-axisymmetric bar potential, the corresponding contributions in these quantities are calculated along ${\phi = 0 \degrees}$ (i.e., the same azimuth as the survey volume). The circular velocity curve is directly related to the radial gravitational forces in the mid-plane; the surface mass density profile within $|z| \leq 1~\text{kpc}$ is a proxy for the vertical gravitational forces following the Jeans equations. The volume density is derived from the (Laplacian of the) potential by means of Poisson's equation. In Fig.~\ref{fig:residualmaps}, $\Sigma_{\mathrm{tot}}$ and $v_{\mathrm{circ}}$ are therefore the physical manifestations of the first derivatives of the potential, while $\rho_{\mathrm{tot}}$ is related to the second derivatives of the potential.
\par
A visual inspection of Fig.~\ref{fig:residualmaps} indicates that for the dataset drawn from the the weak bar model (i.e., \textbf{40m2a01Inter}), the mid-plane volume density, surface density, and the circular velocity is recovered to better than $1$ percent relative error within the radial extent of the survey volume, as well as within the whole radial extent considered here. In other words, in the weak-bar case, \roadmapping\ recovers the underlying \textit{local} and \textit{global} axisymmetric potential and the associated gravitational forces fairly accurately. For the dataset drawn from the the strong bar model (i.e., \textbf{40m2a03Inter}), the recovery of the mid-plane volume density, surface density, and the circular velocity is not as accurate as for the case of a weak bar (compare top and bottom panels of Fig.~\ref{fig:residualmaps}). We have verified that, for the strong bar case, \roadmapping\ can recover -- both \textit{locally} as well as \textit{globally} -- the mid-plane volume density, surface density, and the circular velocity to within $1 \! - \! 2$ percent relative error.
This is especially noteworthy, as \citet{Tricketal2017} found that in the presence of spiral arms, \roadmapping\ recovered the true, spiral-arm-affected, local gravitational forces. Here, in the case of bar resonances, we find that we recover the underlying global axisymmetric potential. 

We have also compared the radial profiles of the volume density, calculated at $z = \pm b_{\rm disc}$, using the true potential $\Phi_{\rm axi}$ and the best-fit potential $\Phi_{\rm axi, fit}$ for both the weak-bar and the strong-bar models. The corresponding volume densities at $z = \pm b_{\rm disc}$ are also recovered within $\sim \! 1 \! - \! 2$ percent relative error, even for the strong bar model (\textbf{40m2a03}). For the sake of brevity, these are not shown here.

\subsubsection{Systematic biases induced by bar resonances}
\label{sec:systematic_paramsearch}

\begin{figure*}
\includegraphics[width=0.95\linewidth]{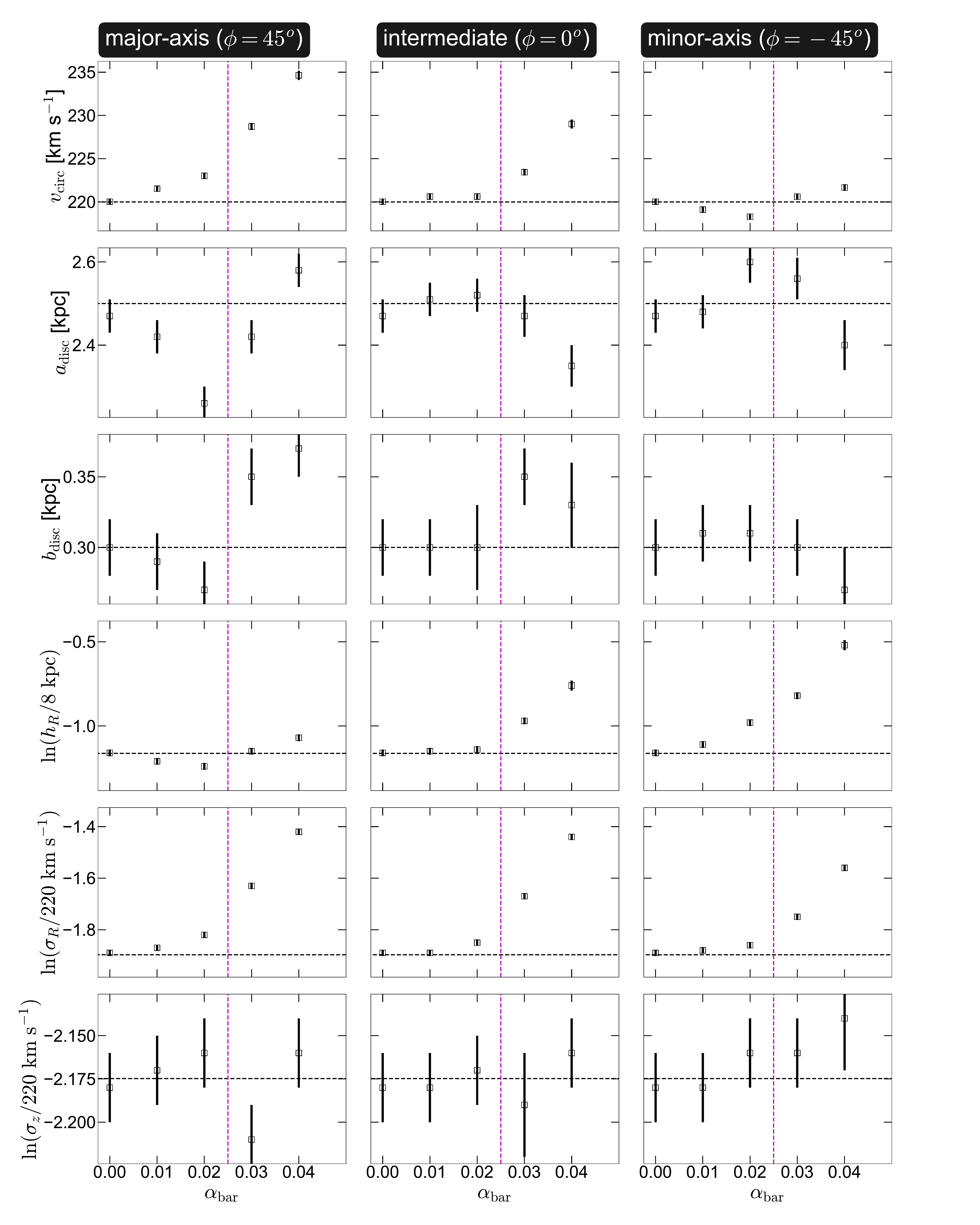}
\caption{Influence of the bar strength and survey-volume location on the recovery of the potential and DF. Shown are the best-fit parameters ($p_{M}$), obtained from the \roadmapping\ analyses applied to the selected survey volumes along the bar's major-axis (left-hand panels), between the major-minor axes (middle panels), and minor-axis (right-hand panels). All survey volumes are centred at ${R_{\rm cen} = 8.2 \kpc}$ and have a radius of ${4 \kpc}$. The bar's pattern speed is ${40~\kmsk}$. The horizontal dashed lines denote the corresponding true value of the parameter in the initial axisymmetric disc. The best-fit parameters deviate systematically from the true values as bar strength ($\alpha_{\rm bar}$) increases. However, the vertical velocity dispersion ($\sigma_{z,0}$) is recovered accurately (within ${1 \sigma}$) in almost all cases. The vertical magenta line denotes ${\alpha_{\rm bar} = 0.025}$, beyond which we see the phenomenon of resonance overlap in our bar models.
}
\label{fig:bestfit_param_finalRuns}
\end{figure*}

Previous sections have revealed the dynamical effect of an $m=2$ bar on the outcome of \roadmapping\ modelling. A bar introduces small but systematic biases in the recovery of the global potential and the DF parameters. Here, we investigate how the systematic biases in the recovery of the global potential and the DF parameters vary with the strength of the bar ($\alpha_{\rm bar}$). We make use of bar models with varying strength (for details, see section~\ref{sec:suite_of_bar_models}) while keeping the azimuthal location of the survey volume (i.e., ${\phi = 0 \degrees}$) unchanged. Fig.~\ref{fig:bestfit_param_finalRuns} shows the resulting best-fit potential and DF parameters obtained by \roadmapping\ as a function of bar strength (see the middle columns of panels there).
\par
As evident from Fig.~\ref{fig:bestfit_param_finalRuns} (middle column of panels), we see a generic trend that the best-fit values of the parameters progressively deviate from their true values with increasing bar strength. For example, the circular velocity, $v_{\rm circ}$ is recovered within $\sim 2.2 \sigma$ for the weakest bar model (\textbf{40m2a01}), whereas for the strongest bar model (\textbf{40m2a04}), the measured $v_{\rm circ}$ differs from the true value by $\sim 18\sigma$. For stronger bar models ($\alpha_{\rm bar} \ge 0.03$), the disc scale length ($a_{\rm disc}$) is systematically under-estimated, thus producing a more centrally concentrated (in the radial direction) best-fit stellar disc. Similarly, the inferred disc scale height ($b_{\rm disc}$), radial tracer density scale length ($h_R$) as well as the radial velocity dispersion ($\sigma_R$) change systematically with increasing bar strength. This is especially true for bar strength $\alpha_{\rm bar} \geq 0.03$, when the bar models begin to display the dynamical phenomenon of resonance overlap. For stronger bars ($\alpha_{\rm bar} \geq 0.03$), the action-space ridge in the ${(J_R,L_z)}$-plane grows larger, and the flattening of the stellar density profile at CR as well as through the net shift of stars towards larger $L_z$ at the OLR become more prominent. This causes a systematic over-estimation of the above-mentioned potential and DF parameters.
Overall, we find that the stronger the bar, the less reliable the formal uncertainties are for our axisymmetric model. This agrees with the intuitive expectation that the stronger the bar is, the less accurately an axisymmetric model will capture the dynamical behaviour of the disc.

The vertical velocity dispersion ($\sigma_{z,0}$) is recovered fairly accurately in almost all cases, within ${1\!  \, \sigma}$. The bar potential we used in this work (see Eq.~\ref{eq:bar_potential_Dehnen}) does not greatly perturb the system in the vertical direction. For more details, see section~5.2 of \citet{Tricketal2021}. Furthermore, we have verified that the increase in the vertical velocity dispersion (as compared to radial velocity dispersion) is negligible in the bar models, relative to the initial axisymmetric model. For the sake brevity, these are not shown here. In addition, the $\sigma_{z,0}$ and $b_{\rm disc}$ parameters remain strongly anti-correlated in all cases, similar to the results shown in sections~\ref{sec:weak_bar} and~\ref{sec:strong_bar}.

\subsection{The influence of the observer location with respect to the bar}
\subsubsection{The azimuthal dependence of resonance signatures}

\begin{figure*}
\includegraphics[width=\linewidth]{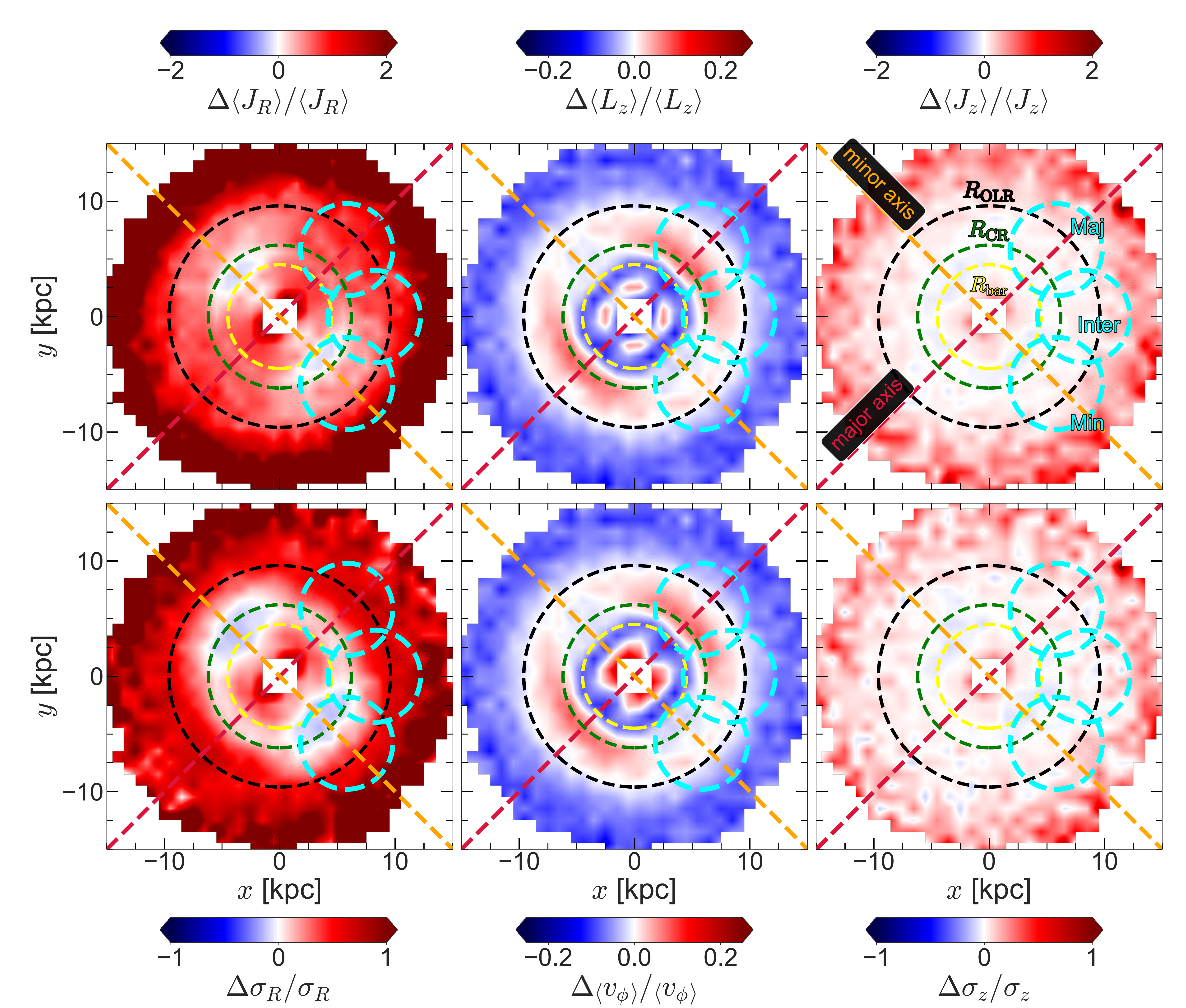}
\caption{Residual distribution of three components of the action ($J_R$, $L_z$, and $J_z$), as well as of the mean azimuthal velocity, and the radial and vertical velocity dispersions, shown in a face-on projection. These residual maps are computed by taking the difference between the strong bar model \textbf{40m2a03} and the initial (true) axisymmetric model. The inner yellow dashed circle denotes the extent of bar (${R_{\rm bar} = 4.5 \kpc}$), while the green and the black dashed circles denote the CR (${R_{\rm CR} = 5.8 \kpc}$) and the OLR (${R_{\rm OLR} = 8.9 \kpc}$) of the bar, respectively. The red and the orange dashed straight lines denote the bar major and the minor axes, respectively. The cyan circles denote the survey volumes used for our \roadmapping\ analyses to obtain the best-fit parameters of the potential and the DF. The large action-space resonant ridge of high $J_R$ values lives across a large range of radii $R$. Even though the low-$J_R$, high-$L_z$ tail of the initial axisymmetric disc is not modified by the bar, the ridge causes an apparent orbital heating (in $J_R$ and $\sigma_R$), and therefore an average decrease of $L_z$ and $v_\phi$ in the outer regions of the discs. 
}
\label{fig:radialheating_bar}
\end{figure*}

Stellar bars are known to heat up disc orbits (see references in section~\ref{sec:Intro}) and to introduce strong non-circular streaming motions \citep[e.g. see][]{Randriamampandryetal2016}. These dynamical phenomena are known to become more prominent as bar strength increases, and as we saw in the previous sections, they can affect the outcome of axisymmetric dynamical modelling. To gain a deeper understanding of the physical reasons behind the observed systematic biases in recovered potential and DF parameters, we measure changes caused by the bar in the face-on distributions of the actions ($J_R$, $L_z$, and $J_z$), as well as of the mean azimuthal velocity, the radial and the vertical velocity dispersion. First, we measure the means of these quantities in pixels in the ${(x,y)}$-plane for the initial axisymmetric model and then for one strong bar model, \textbf{40m2a03}. Then, to measure the change quantitatively, we calculate residual maps for these quantities, obtained by taking the difference between the bar model and the axisymmetric model. These are shown in Fig.~\ref{fig:radialheating_bar}.
%
%
\par
As seen in Fig.~\ref{fig:radialheating_bar}, radial heating of stellar orbits (caused by the bar) is manifested as increments in radial actions ($J_R$) and radial velocity dispersions ($\sigma_R$). Interestingly, the increase in the radial velocity dispersion as well as radial action (both can be considered as proxy for the radial heating) show a characteristic azimuthal variation at fixed radius; the stars along the bar major axis are preferentially heated to a larger degree when compared with the stars along the bar minor axis. Using a suite of self-consistent $N$-body models of barred galaxies, recent work by \citet{Ghoshetal2022a} also demonstrated azimuthal variation of the heating of stars by the stellar bar (see Fig.~9 there).
In addition, the changes in average $L_z$, especially near the corotation, are caused by the bar-induced inward/outward radial migration of the stars along the major axis. The vertical action ($J_z$) as well as the vertical velocity dispersion ($\sigma_z$) do not display much variation (as compared to their in-plane counterparts) with the introduction of the bar. This implies that the dynamical effect of the bar in our models remains mostly limited in the in-plane orbits.
Motivated by these trends, we next investigate the dependence of \roadmapping{}'s recovered potential and DF parameters on the azimuthal location of the survey volume considered.

\subsubsection{Systematic modelling biases depending on observer location}
\label{sec:bias_with_azimuth}
Here, we consider three survey volumes for each of the bar models, along the bar major axis, minor axis, and in-between (for details, see section~\ref{sec:survey_volume}). Then, for each of these 12 selected datasets, we perform the \roadmapping\ analysis. The corresponding best-fit values of the parameters and their associated errors as a function of azimuthal location and bar strength ($\alpha_{\rm bar}$) are shown in Fig.~\ref{fig:bestfit_param_finalRuns}. Overall, we see a generic trend that the best-fit values of the parameters depart from their true values for stronger bar models, irrespective of the azimuthal location (w.r.t the bar) of the survey volume. However, for a fixed value of the bar strength ($\alpha_{\rm bar}$), potential and DF parameters are, in general, better recovered for survey volumes lying along the bar minor-axis. For most of the cases, the largest biases (due to the bar) are found for survey volumes lying along the bar major axis. This is consistent with the $\phi$-variation of the resonance signatures at OLR and CR observed in Fig.~\ref{fig:radialheating_bar}. Furthermore, for bar strengths $\alpha_{\rm bar} \gtrsim 0.025$, the bar models begin to display the dynamical phenomenon of resonance overlap, thereby introducing even larger and also qualitatively different systematic bias trends in the best-fit parameters than for $\alpha_{\rm bar} < 0.025$ (see Fig.~\ref{fig:bestfit_param_finalRuns}).
Below we explain a few trends uncovered in this systematic analysis.

\begin{itemize}
\item{For bar models with resonance overlap (i.e., for $\alpha_{\rm bar} \geq 0.03$), the rotational velocity ($v_{\rm circ}$) is systematically over-estimated. This over-estimation is greater along the bar major axis than along the bar minor axis. As is evident from Fig.~\ref{fig:radialheating_bar}, mean azimuthal velocity, $\avg{v_{\phi}}$, increases to a greater extent along the bar major axis than along the bar-minor axis. In the presence of resonance overlap, $v_{\rm circ}$ is therefore over-estimated to a greater degree in survey volumes along the bar major axis. For bar models without resonance overlap (i.e., for $\alpha_{\rm bar} < 0.03$), $v_{\rm circ}$ is under-estimated for survey volumes along the bar minor axis, but over-estimated for survey volumes along the bar major axis. We have verified that for the bar model with $\alpha_{\rm bar} = 0.02$, there is a net increase in $\avg{v_{\phi}}$ along the bar major axis, while $\avg{v_{\phi}}$ decreases along the bar minor axis. This explains the trend of over- and under-estimation in $v_{\rm circ}$ along bar major and minor axes, respectively.
}

\item{With increasing bar strength ($\alpha_{\rm bar}$), the radial velocity dispersion, $\sigma_R$, deviates systematically from the value corresponding to an axisymmetric configuration. The OLR radially heats the orbits by creating a ridge, and this heating process increases with the bar strength; thereby leading to increasing departure of the model from its initial axisymmetric configuration. This, in turn, leads to increasing deviation of $\sigma_R$. Furthermore, for stronger bars ($\alpha_{\rm bar} \geq 0.02$), $\sigma_R$ is more accurately recovered along the bar minor axis than along the bar minor axis. We have verified that the OLR heats stars to a lesser extent along the bar minor axis. Furthermore, in bar models with resonance overlap, the OLR ridge is stronger along the bar major axis. As a consequence, $\sigma_R$ is more accurately recovered along the bar minor axis.
}
\item{For bar models with resonance overlap (i.e., for $\alpha_{\rm bar} \geq 0.03$), $h_R$ also systematically deviates from its initial value (corresponding to an axisymmetric configuration). On average, the OLR moves stars from inside the OLR to outside of the OLR. In the case of resonance overlap, the 1:1 resonance moves stars even further outwards. This `inverts' the radial density gradient around the OLR and therefore increases the measured $h_R$. In addition, we find that the bias is stronger along the bar minor axis. Furthermore, for bar models without resonance overlap (i.e., for $\alpha_{\rm bar} < 0.03$), $h_R$ is under-estimated along the bar major axis, and over-estimated along the bar minor axis.
}
\end{itemize}

\subsection{The influence of pattern speed and disc coverage}
\label{sec:pattern_speed_and_disc_coverage}

In the previous sections, we investigated spherical survey volumes with radius $d_{\rm max} = 4 \kpc$  and applied \roadmapping\ to recover the parameters of the potential as well as the DF. However, constructing tracer datasets in the real Milky Way that are volume complete in the presence of dust extinction and that also have precise 6-D measurements as far out as $4~\text{kpc}$ from the Sun will be a challenge, even with \gaia\ data. It is therefore important to investigate whether the success of the recovery of these parameters also depends on the size of the chosen survey volume in the Galactic disc. In a perfectly axisymmetric disc, even a very small survey volume does not lead to systematic biases in the \roadmapping\ recovery \citep{Tricketal2016}. However, we find that in the case of barred models, smaller survey volumes are more dominated by the non-axisymmetric resonances, as illustrated in Fig.~\ref{fig:action_space_comparison_between_pattern_speed_and_survey_volume_size}.
To test the effect of survey volume on \roadmapping{}, we first choose the case \textbf{40m2a03Inter}, and then vary the radius of the survey volume $d_{\rm max}$ between $4, 3,$ and $2 \kpc$. 
Furthermore, depending on the exact value of the pattern speed of the MW bar, the Solar neighbourhood might also be placed differently with respect to the bar resonances, as can be seen in Fig.~\ref{fig:action_space_comparison_between_pattern_speed_and_survey_volume_size} (compare the upper and lower panel). To mimic such an effect, we also consider a bar model \textbf{28m2a03} (for details, see section~\ref{sec:suite_of_bar_models}), and then choose survey volumes along ${\phi = 0 \degrees}$ while varying the $d_{\rm max}$ between $4, 3,$ and $2 \kpc$. 
Fig.~\ref{fig:effect_of_survey_Volume} shows the corresponding best-fit parameter values from \roadmapping{} for both simulations with different pattern speeds, as function of $d_{\rm max}$.

\subsubsection{Pattern speed}
\label{sec:pattern_speed_variation}
\begin{figure}
    \centering\includegraphics[width=0.7\columnwidth]{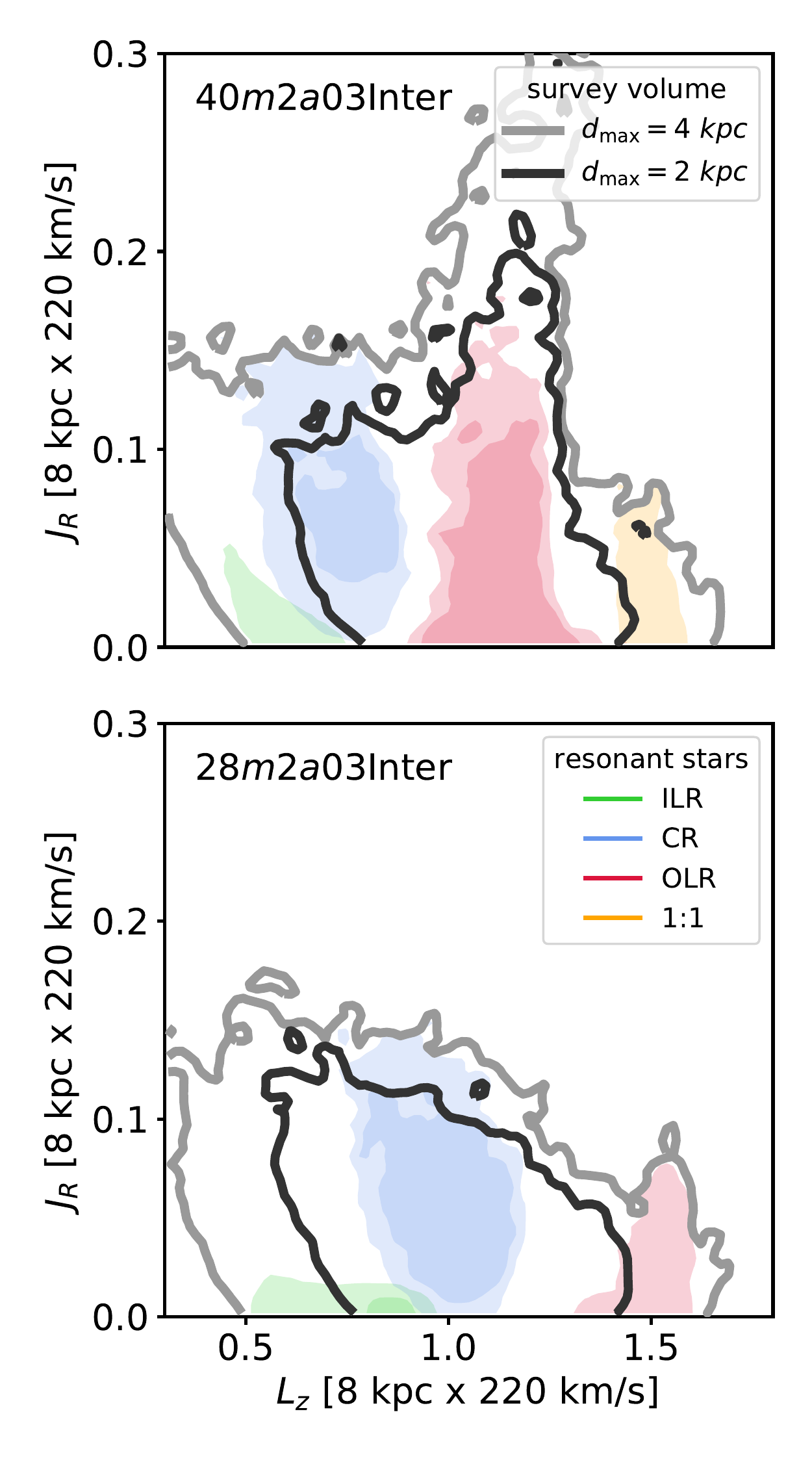}
    \caption{Comparison of the action distribution for the two bar models, with pattern speeds ${40 \kmsk}$ and ${28 \kmsk}$, and two of the survey volumes (${d_\text{max}=2~\text{kpc}}$ and ${4~\text{kpc}}$) at ${\phi_{\text{cen}} = 0 \degrees}$ analysed with \roadmapping\ in Fig.~\ref{fig:effect_of_survey_Volume}. The contours are all drawn at a level of 1 particle, indicating the outer edges of the action distribution. The thick grey smoothed contours show the distribution for all stars in the survey volumes, and the coloured filled contours those of the stars in resonance with the bar according to Eq.~\eqref{eq:true_resonance_condition}. These four cases demonstrate scenarios in which different resonances dominate the dataset to different extents.}
    \label{fig:action_space_comparison_between_pattern_speed_and_survey_volume_size}
\end{figure}

As seen in Fig.~\ref{fig:effect_of_survey_Volume}, for a fixed value of $d_{\rm max}$, the best-fit parameters are almost always more accurately recovered for the \textbf{28m2a03} model than for the \textbf{40m2a03} model. Although the bar strengths at $R_{\odot}$ are the same in these models, for the model \textbf{28m2a03}, the resonances lie further out in galactocentric radius, and are therefore slightly weaker. This is enough that for a similarly strong bar, resonance overlap does not occur in the model \textbf{28m2a03}. This allows \roadmapping\ to smooth over the resonances and the regions unaffected by the bar. \roadmapping\ therefore recovers the best-fit parameters more accurately for \textbf{28m2a03} (compare red and black points in Fig.~\ref{fig:effect_of_survey_Volume}). For both pattern speeds, stars in resonance with CR dominate the dataset close to the observer, leading to over-estimation of $v_{\rm circ}$ and $h_R$ (see Fig.~\ref{fig:action_space_comparison_between_pattern_speed_and_survey_volume_size} and discussions in section~\ref{sec:bias_with_azimuth}). For the model \textbf{28m2a03}, the OLR plays almost no role, and $\sigma_R$ is therefore almost perfectly recovered, as compared to the model \textbf{40m2a03}, which is dominated by the OLR ridge (see Fig.~\ref{fig:action_space_comparison_between_pattern_speed_and_survey_volume_size}).

\subsubsection{Size of the survey volume}
\label{sec:survey_volume_effect}

%
%
\begin{figure*}
\includegraphics[width=\linewidth]{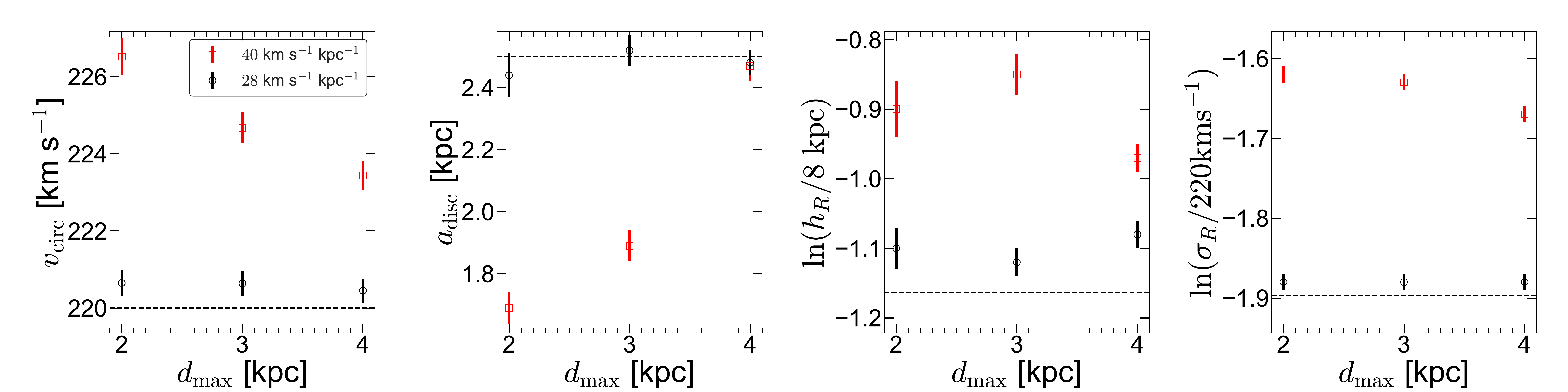}
\caption{
Influence of the survey volume size on the recovery of potential and DF parameters. Shown are the best-fit parameters ($p_{M}$), obtained by applying \roadmapping\ to the given survey volumes, all selected along ${\phi = 0 \degrees}$ (`Inter' survey volumes) but having different spatial extents (i.e., varying radius $d_{\rm max}$ from the observer, located at ${R = 8.2 \kpc}$). The bar strength is ${\alpha_{\rm bar} = 0.03}$, and two bar pattern speeds, namely, ${40 \kmsk}$ and ${28 \kmsk}$ are used (see legend). The horizontal dashed lines denote the corresponding true parameter values. With smaller survey volumes, the best-fit values of the parameters progressively drift away from their respective true values. This is particularly true for the potential parameters ($p_{\Phi}$). The systematic bias introduced by the bar in the recovered disc scale length $a_\text{disc}$ is substantial, even for a survey volume as large as ${d_{max}=3~\text{kpc}}$. This emphasizes that axisymmetric dynamical modelling works best if a large region in the Galaxy is probed, allowing the model to average over non-axisymmetric structures.
}
\label{fig:effect_of_survey_Volume}
\end{figure*}
\par
As seen in Fig.~\ref{fig:effect_of_survey_Volume}, as the survey volume shrinks, the best-fit values of the parameters progressively drift away from their respective true values. This trend is more evident for survey volumes taken from the bar model \textbf{40m2a03}. A similar trend is also seen for the \textbf{28m2a03} model (albeit with lesser variation with $d_{\rm max}$). The larger the survey volume, the better \roadmapping\ can average over the resonances and recover the potential. This is best seen for the disc scale length ($a_{\rm disc}$). For the circular velocity ($v_{\rm circ}$), the systematic biases do not vanish completely with increasing $d_{\rm max}$, but are small in absolute terms. In addition, we find that the DF parameters are less sensitive to survey volume size and remain systematically biased. This is due to the dominance of stars in resonance with the CR and OLR within the chosen survey volumes (for more detail, see sections~\ref{sec:recap_resonances_action_space} and~\ref{sec:recover_df}).
This further indicates that constraining the potential is not at all hopeless in the presence of a bar (in case of a large volume and/or in the absence of resonance overlap).
 
\section{Discussion}
\label{sec:discussion}
\subsection{Caveats}

In this work, we use a suite of test-particle simulations of barred disc galaxies with varying bar properties (pattern speed, strength). 
While this serves our purpose of testing the applicability and success of action-based dynamical modelling in the presence of non-axisymmetric features, such as a stellar bar, these test-particle simulations are not self-consistent, and the evolution and growth of the stellar bar is not tracked in a self-consistent fashion. We plan to investigate this in future work using disc galaxies from cosmological zoom-in simulations.
Furthermore, the MW has both a bar and spiral arms in the disc (see references in section~\ref{sec:Intro}). While \citet{Tricketal2017} investigated possible biases introduced due to the presence of spiral arms, and in this work we have investigated the same issue for an ${m=2}$ bar, the joint dynamical effect of a bar and spirals on action-based dynamical modelling of MW-like galaxies nevertheless remains to be investigated. We also plan to address this question in future work.
\par
Secondly, the mock data (drawn from the chosen survey volumes) from our bar models is \textit{perfect}. While our mock data is error-free, $6-$D phase-space information from any real observations would contain associated measurement uncertainties/errors. \citet{Tricketal2016} dealt with this issue in great depth in the context of the \roadmapping\ framework. Here, we stress that the main focus of this work has been to identify and quantify the dynamical effect of the non-axisymmetric bar feature on action-based dynamical modelling. Hence, we do not include any measurement uncertainties/errors in our mock datasets.
\par
Lastly, we point out that the mock data-sets (drawn from different survey volumes) used in this paper are volume-complete. However, in general, completeness decreases with distance from the observer. This implies that one needs to make sure to either consider stellar sub-populations that are volume-complete (e.g. standard candles, such as red clump stars), or to more carefully model the selection function and to additionally make sure that the number of stars included at a distance of ${4 \kpc}$ is still large enough to have a meaningful effect on the fit.

\subsection{Implications for modelling the Milky Way}
\label{Implications_for_MW}
Overall, we find that for a large survey volume ($\sim 4 \kpc$), and along an azimuth falling in-between the bar major and minor axes, the recovery of the local potential (Fig.~\ref{fig:residualmaps}) and of the potential parameters (Fig.~\ref{fig:bestfit_param_finalRuns}) is not catastrophic, even in the presence of a strong bar with ${\alpha_{\rm bar} = 0.03}$. In fact, in this case, we still recover two of the potential parameters ($v_{\rm circ}$ and $a_{\rm disc}$) to within $1-2$ percent of their true values, and the third potential parameter, $b_{\rm disc}$, to within $\sim \! 17$ percent (for details, see section~\ref{sec:strong_bar}). This is a promising trend as far as the recovery of the potential of the MW is concerned. As demonstrated in Fig.~\ref{fig:bestfit_param_finalRuns}, if the observer is close to the major axis of the bar, we expect stronger biases in the recovered potential parameters than for an observer along the minor axis. The MW's bar is oriented at ${\sim \! 28 \! - \! 33 \degrees}$ with respect to the Solar neighbourhood \citep[e.g.,][]{Weggetal2015}, and in light of our findings here, this makes the possibility of recovering the MW's potential somewhat difficult. In addition, in the near future, we do not expect that we will be able to construct volume complete datasets with precise 6D data further out than $2 \kpc$ from the Sun. Dust obscuration and the apparent magnitude limit of the \gaia\ RVS become dominant as one moves further out, particularly in the inner Galaxy. In order to construct MW samples that extend to greater distances, it becomes increasingly important to accurately model the dust-affected selection function \citep{Bovyetal2016}.
If we are limited to smaller survey volumes, the ability of \roadmapping\ (or action-based dynamical modelling in general) to constrain the MW potential depends on the combination of the bar pattern speed and strength. If there is an overlap of the bar resonances, \roadmapping\ will not have not enough `close-to-axisymmetric' regions in the data to average over the resonances and provide an accurate estimate of the potential. The slow bar (${\Omega_{\rm bar} = 39 \kmsk}$, ${\alpha_{\rm bar} = 0.024}$) scenario, as used in \citet{Portailetal2017,2017ApJ...840L...2P,2019A&A...626A..41M}, which associates the Hat moving group with the OLR and Hercules with CR, does not create the kind of large resonance-overlap ridges we see in the simulations in this work. In addition, as illustrated in Fig.~\ref{fig:effect_of_survey_Volume}, even if the bar were as strong as ${\alpha_{\rm bar} = 0.03}$, we would still be able to constrain the potential quite well in a slow-bar scenario. For an intermediate bar with ${\Omega_{\rm bar}= 41 \kmsk}$, we have less leeway in terms of bar strength (see Fig.~\ref{fig:bestfit_param_finalRuns}). However, given that all dominant ridges in the \gaia\ data have negative slopes \citep{Tricketal2019,Tricketal2021}, and our resonance-overlap ridge in this work has a positive slope, we are optimistic that this kind of strong resonance overlap does not occur in the MW. An intermediate bar in the MW would would be weaker than ${\alpha_{\rm bar} = 0.03}$, and would therefore providing a good basis for reliable \roadmapping\ potential constraints.

\section{Summary}
\label{sec:summary}

In this work, we have investigated how the presence of a bar affects the outcome of action-based dynamical modelling of disc galaxies. The classical action-based modelling approach assumes axisymmetry and phase-mixing, and does not account for bar resonances that are known to be present in, for example, our MW's disc. To examine the systematic biases this introduces, we made use of \roadmapping{}---a well-tested framework for action-based dynamical modelling---and applied it to a set of test-particle models of barred galaxies while varying the properties of the stellar bar. Our main findings are:\\

\begin{itemize}
\item{We demonstrate that even though bars can have a strong effect on the DF of the stars orbiting in a MW-like disc, action-based dynamical modelling still works in presence of bar resonances. For realistic bar parameters ($\alpha_{\rm bar} = 0.03$, $\Omega_{\rm bar} = 40 \kmsk$), the global potential parameters are recovered to within ${\sim \! 1 \! - \! 17}$ percent. However, the DF parameters, especially those controlling the shape of the DF in the radial direction, are significantly affected due to the combined effects of radial heating, resonance overlap, and migration induced by bar resonances.
}
\item{We find that the recovery of the potential and the DF parameters using action-based dynamical modelling depends on the azimuthal position of the chosen survey volume with respect to the bar major axis. For survey volumes lying along the bar major axis, the recovery of the global potential and DF parameters is significantly affected for stronger bars. This is because a stronger bar radially heats the disc stars, preferentially along the bar major axis, thereby inducing a larger deviation from the axisymmetric configuration assumed by the model.
}

\item{The size and the location of the survey volume (with respect to the resonances) play a pivotal role in the success of the recovery of the potential and the DF parameters using \roadmapping{}. The potential parameters are better constrained for larger survey volumes. However, the DF parameters are less sensitive to the size of the survey volume, and remain systematically biased due to the dominance (within the survey volumes) of stars in resonance with CR and OLR. Similarly, if the survey volume lies farther from the resonances (CR and OLR), so that the stars in the survey volume is not dominated by resonant stars, the potential and the DF parameters are better recovered by \roadmapping{}.  }

\end{itemize}

To conclude, given the position of the Solar neighbourhood with respect to the MW's bar, and the availability of the rich stellar $6-$D phase-space information from the recent \gaia\ DR3, in the Solar neighbourhood and beyond, it is still possible to use action-based dynamical modelling to estimate the underlying axisymmetric potential (and the associated DF) for the MW with good accuracy, even in presence of the bar.

\section*{Acknowledgement}
We thank the referee, Paul McMillan, for useful comments which helped to improve this paper.
S.G. and G.M.G. acknowledge funding from the Alexander von Humboldt Foundation, through the Sofja Kovalevskaja
Award. The simulations and the \roadmapping\ analysis are performed on the GPU computing facility at the Max-Planck-Institut f\"{u}r Astronomie (MPIA), Heidelberg, Germany.

\section*{Data Availability}
\noindent
All of the test-particle simulations of barred galaxies used in this work are publicly available at \url{https://doi.org/10.5281/zenodo.7413918}.

\bibliography{my_ref}{}
\bibliographystyle{mnras}


\appendix

\section{Formation of ridge in the action-space and resonance overlap}
\label{appen_ridgeFormation}
\begin{figure*}
\includegraphics[width=0.95\linewidth]{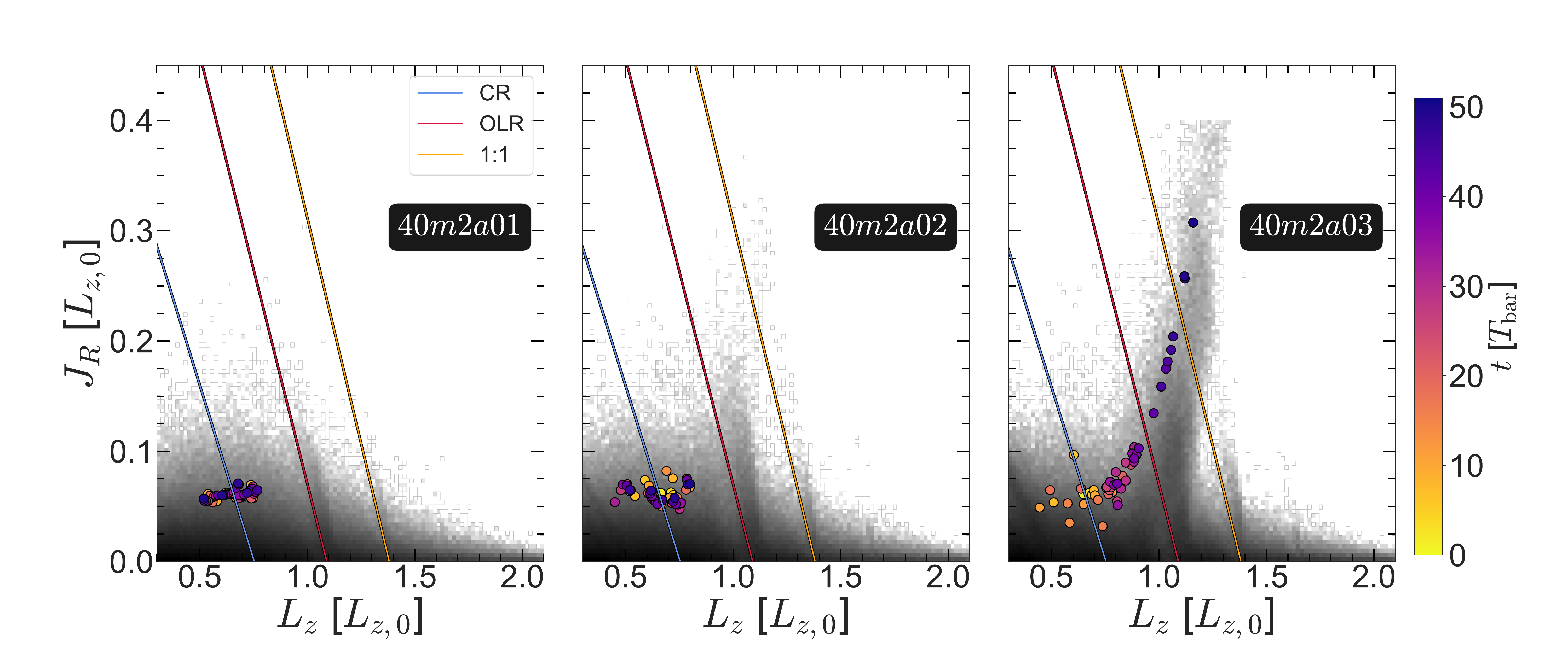}
\caption{Trajectory of a stellar particle, obtained by integrating its orbit under the influence of the joint potential ($\Phi_{\rm axi} + \Phi_{\rm bar}$) while varying the bar strength ($\alpha_{\rm bar}$), shown as a function of time (see the colourbar). The grey filled contours denote the overall stellar distribution of the corresponding bar model. In each panel, the straight lines denote the corresponding ARLs (for details see section~\ref{sec:recap_resonances_action_space}). As throughout this work, ${L_{z,0} = 8 \times 220 \kkms}$. For bar strength $\alpha_{\rm bar} \leq 0.02$, the chosen stellar particle's libration is limited to the CR. However, for the strong bar model, \textbf{40m2a03}, the slope of the libration becomes steeper when moving from CR to the 1:2 OLR to the 1:1 resonance, and the stellar particle finally ends up in the strong-ridge region.}
\label{fig:appen_ridgeform}
\end{figure*}

Earlier in section~\ref{sec:recap_resonances_action_space}, we showed that for bar strengths of $\alpha_{\rm bar} \geq 0.03$, our barred models display a strong ridge in the ${(J_R,L_z)}$-plane. We attributed the formation of this strong ridge to scattering at the 1:2 OLR, and then to the phenomenon of resonance overlap. Here, we investigate this in more detail. We first choose the strong bar model, namely, \textbf{40m2a03}, and randomly select 100 stars lying on the ridge at $t= 50 T_{\rm bar}$. Then, we identify these selected stars at the beginning of the simulation run ($t = 0$). We integrate the orbits of these selected stars for the same time-span ($t= 50 T_{\rm bar}$) under the influence of the joint potential ($\Phi_{\rm axi} + \Phi_{\rm bar}$) for different bar strengths ($\alpha_{\rm bar}$). One such example of the particle's excursion in the ${(J_R,L_z)}$-plane, as a function of time, is shown in Fig.~\ref{fig:appen_ridgeform}. As seen from Fig.~\ref{fig:appen_ridgeform}, when the bar is weak ($\alpha_{\rm bar} \leq 0.02)$, the particle's libration remains limited to regions around the CR. For these models, resonance overlap does not occur, and a strong ridge does not form. However, for the strong-bar model, \textbf{40m2a03}, the slope of the libration changes from $\Delta J_R / \Delta L_z = l/m = 0$ for the CR to $l/m = 1/2$ for the OLR, and to $l/m = 1/1$ for the 1:1 resonance (i.e., the slope of the libration becomes steeper when moving from CR to the OLR to the 1:1 resonance), and the stellar particle finally ends up in the strong-ridge region. This demonstrates the vital role of scattering at the 1:2 OLR and the resonance overlap for strong bar models in creating strong ridge in the ${(J_R,L_z)}$-plane.

\bsp
\label{lastpage}

\end{document}